\documentclass[%
aip,
jcp,%
amsmath,amssymb,
reprint,%
longbibliography
]{revtex4-1}
\usepackage[latin1]{inputenc}

\usepackage{amsmath}%
\usepackage{amsfonts}
\usepackage{amssymb,bm}
\usepackage{graphicx}
\usepackage{physics}
\usepackage{upgreek}
\usepackage[outercaption]{sidecap}   
\usepackage{color}
\usepackage[normalem]{ulem}
\mathchardef\mhyphen="2D
\usepackage{txfonts}

\usepackage[nice]{nicefrac}

\DeclareMathAlphabet{\pazocal}{OMS}{zplm}{m}{n}

\date{\today}

\begin{document}

\def\bra#1{\left<{#1}\right|}
\def\ket#1{\left|{#1}\right>}
\def\expval#1#2{\bra{#2} {#1} \ket{#2}}
\def\mapright#1{\smash{\mathop{\longrightarrow}\limits^{_{_{\phantom{X}}}{#1}_{_{\phantom{X}}}}}}

\title{\vspace{0.0cm}  Semiclassical instanton theory for reaction rates at any temperature: How a rigorous real-time derivation solves the crossover temperature problem}

\author{Joseph E.\ Lawrence}
\affiliation{\mbox{Simons Center for Computational Physical Chemistry, New York University, New York, NY 10003, USA}
\\
Department of Chemistry, New York University, New York, NY 10003, USA}
\email{joseph.lawrence@nyu.edu}

\begin{abstract}
Instanton theory relates the rate constant for tunneling through a barrier to the periodic classical trajectory on the upturned potential energy surface whose period is $\tau=\hbar/(k_{\rm B}T)$. 
Unfortunately, the standard theory is only applicable below the ``crossover temperature'', where the periodic orbit first appears.
This paper presents a rigorous semiclassical ($\hbar\to0$) theory for the rate that is valid at any temperature.  The theory is derived by combining Bleistein's method for generating uniform asymptotic expansions with a real-time modification of Richardson's flux-correlation function derivation of instanton theory.
The resulting theory smoothly connects the instanton result at low temperature to the parabolic correction to Eyring transition state theory at high-temperature. 
Although the derivation involves real time, the final theory only involves imaginary-time (thermal) properties, consistent with the standard version of instanton theory.
Therefore, it is no more difficult to compute than the standard theory.
The theory is illustrated with application to model systems, where it is shown to give excellent numerical results.
Finally, the first-principles approach taken here results in a number of advantages over previous attempts to extend the imaginary free-energy formulation of instanton theory. In addition to producing a theory that is a smooth (continuously differentiable) function of temperature, the derivation also naturally incorporates hyperasymptotic (i.e.~multi-orbit) terms, and provides a framework for further extensions of the theory. 

\end{abstract}

\maketitle

\section{Introduction}
The cornerstone of chemical reaction rate theory is Eyring transition state theory (TST).\cite{Eyring1935} Developed in the 1930s,\cite{Eyring1935,Eyring1938rate,Wigner1932parabolic,Wigner1937TST,Evans1935TST,Miller1998review} TST is still widely used today to estimate the rate of chemical reactions.\footnote{See W.~H. Miller, Faraday Discuss. {\bf 110}, 1-21 (1998) for a detailed historical account of the development of transition state theory.} The advantage of the theory is its simplicity, allowing for the estimation of the rate constant from only local knowledge of the Born-Oppenheimer potential in the reactant minimum and at the saddle point separating the reactants and products. The TST approximation to the rate has the simple form
\begin{equation}
    k_{\rm TST}  =  \frac{\kappa}{2\pi\beta\hbar} \frac{Z^\ddagger}{Z_r} \, e^{-\beta V^\ddagger}
\end{equation}
where $\beta=1/k_{\rm B} T$ is the inverse temperature, $V^\ddagger$ is the height of the potential barrier, $\kappa$ is the transmission coefficient, $Z_r$ is the reactant partition function and $Z^\ddagger$ is the transition state partition function for the modes orthogonal to the unstable coordinate. Typically, both $Z_r$ and $Z^\ddagger$ are treated quantum mechanically, using a harmonic approximation for the vibrations and rigid-rotors for the rotations. In contrast, in the basic version of the theory, motion over the barrier is assumed to be classical, which in the absence of dynamical recrossing leads to a transmission coefficient of $\kappa=1$. 

Already in the 1930s, it was clear to the developers of TST that quantum tunneling along the reaction coordinate could result in a transmission coefficient, $\kappa>1$.
When the effect of tunneling is small, a reasonable approximation is given by Wigner's famous tunneling correction\cite{Wigner1932parabolic}
\begin{equation}
    \kappa_{\rm W}  = 1 + \frac{(\beta \hbar \omega)^2}{24},
\end{equation}
where $\omega$ is the barrier frequency along the unstable coordinate. This is, in fact, just the first term in an expansion of the exact tunneling correction for the parabolic barrier\cite{BellBook}     
\begin{equation}
    \kappa_{\rm PB} = \frac{\beta \hbar \omega /2  }{ \sin(\beta\hbar\omega/2)}.
\end{equation}
Nevertheless, $\kappa_{\rm W}$ is typically preferred for practical purposes over $\kappa_{\rm PB}$ as the parabolic barrier approximation diverges at the  ``crossover temperature'' $ T_{\rm c}=\frac{\hbar \omega}{2\pi k_{\rm B}}$.
At low temperatures, where tunneling can enhance the rate by many orders of magnitude, such simple corrections are insufficient. Therefore, a commonly used approach is to retain the separable approximation of TST and compute a one-dimensional tunneling correction, e.g.~by fitting the barrier to a function for which the result is known analytically,\cite{Miller1979unimolecular} or by using a semiclassical tunneling probability based on Wentzel--Kramers--Brillouin (WKB) theory.\cite{BellBook} These one-dimensional approximations, however, fail to capture non-separable effects such as ``corner cutting'', where the system tunnels through higher but narrower regions of the potential.\cite{Marcus1977cornercut,FernandezRamos2007review} 

Instanton theory provides a rigorous way to go beyond these simple one-dimensional tunneling corrections. The instanton is the dominant tunneling path in a full-dimensional path-integral description of the reaction.\cite{Richardson2018InstReview}
Hence, it is not constrained to follow the minimum energy path and can, therefore, capture key nonseparable 
effects on the rate.
In fact, rather than being a minimum energy path, the instanton is a stationary action path. Following Lagrange's principle, the instanton can thus be interpreted as a classical trajectory. Specifically, it is a periodic imaginary-time trajectory (which is equivalent to a real-time trajectory on the upturned potential) whose period is the thermal time, $\tau=\beta\hbar$. The instanton can be found practically by optimisation of the action for a discretised path,\cite{Andersson2009Hmethane,Richardson2009RPInst}
which is not significantly more computationally challenging than locating the transition state.\cite{GPR,Fang2024CheapInstantons} 
This approach has been used to apply instanton theory to a wide range of systems in full dimensions, both with pre-computed potential energy surfaces and also on-the-fly using high-level electronic structure theory.\cite{GPR,Fang2024CheapInstantons,HCH4,Meisner2011isotope,Cooper2018interpolation,McConnell2019instanton,Litman2022InstantonElectronicFrictionII}
Having located the instanton path, the instanton approximation to the rate constant can be written in a similar form to Eyring-TST\cite{Miller1975semiclassical}
\begin{equation}
    k_{\rm inst} =  \frac{1}{\sqrt{2\pi\hbar}} \left(- \frac{\mathrm{d}^2 S_{\!\rm inst}}{\mathrm{d}\tau^2} \right)^{\nicefrac{1}{2}} \frac{Z_{\rm inst}}{Z_r} e^{-S_{\!\rm inst}(\tau)/\hbar} \label{eq:Standard_Instanton}
\end{equation}
where $S_{\!\rm inst}(\tau)$ is the instanton action and $Z_{\rm inst}$ the instanton partition function. Here, $Z_{\rm inst}$ generalises $Z^\ddagger$ by capturing the effect of the changing vibrational frequencies and rotational constants along the instanton path. This can be seen explicitly in the formal expression for $Z_{\rm inst}$, which in the absence of rotations is\cite{Miller1975semiclassical}
\begin{equation}
    Z_{\rm inst}(\tau) = \prod_{j=1}^{f-1} \frac{1}{2\sinh[u_j(\tau)/2]},
\end{equation}
here $u_j(\tau)$ is the $j^{\,\rm th}$ stability parameter for the instanton orbit, which for a separable system reduces to $u_j(\tau)=\tau\omega_j^{\rm vib}$.

Part of the power of instanton theory is that it is a rigorous semiclassical theory.\cite{Richardson2016FirstPrinciples,Richardson2018InstReview} 
In particular instanton theory is the first term in an asymptotic series expansion of the exact quantum rate as $\hbar\to0$. 
Asymptotic expansions can be thought of as generalising the idea of a perturbative expansion to problems where a simple power series may not be applicable.\cite{BenderBook,RWong1989Book} An important example is obtaining the expansion in terms of $\varepsilon$ of an integral of the form
\begin{equation}
    I(\varepsilon) = \int_{-\infty}^\infty A(x) \, e^{-f(x)/\varepsilon} \, \mathrm{d}x  
\end{equation}
for $\varepsilon\to0$. The asymptotic expansion for this integral can be found by Laplace's method (or equivalently steepest descent integration), resulting in an asymptotic series of the form\cite{BenderBook,RWong1989Book}$^{,}$\footnote{Where it is assumed $A(x^\star)\neq0$.}  
\begin{equation}
    I(\varepsilon)  \sim A(x^\star)\sqrt{\frac{2\pi\varepsilon}{f''(x^\star)}} \, e^{-f(x^\star)/\varepsilon} ( 1+ a_1 \varepsilon + a_2 \varepsilon^2 + \dots ), \label{eq:steepest_descent}
\end{equation}
where $x^\star$ is the global minimum of $f(x)$, i.e.~$f'(x^\star)=0$. We see that, at leading order, this corresponds to approximating the integrand by expanding $f(x)$ to second order about $x^\star$ and treating $A(x)$ as constant. 
The apparent simplicity of this procedure belies the power and rich complexity of asymptotic analysis. 
For example, although asymptotic series are generally not convergent, their first few terms typically give a very good approximation to the exact result. When both representations are available, asymptotic series are usually much more numerically efficient than a Taylor series representation.\cite{Dingle1973AsymptoticsBook}
% and when both are available are usually much more numerically efficient than a Taylor series representation. 
Furthermore, despite the apparently approximate nature of asymptotics, in principle a thorough asymptotic analysis in combination with modern resummation methods allows exact results to be recovered.\cite{Dingle1973AsymptoticsBook,Ecalle1981Resurgence,Dunne2014ResurgentTransseries}  

One might at this stage reasonably ask, what one means by an expansion in $\hbar$? In particular, there is an apparent ambiguity in whether one defines the inverse temperature in terms of the thermal time, $\beta=\tau/\hbar$, or vice-versa $\tau=\beta\hbar$. 
The ambiguity is resolved by a more formal definition, such that, $\hbar\to0$ is actually shorthand for introducing a perturbation parameter $\varepsilon$ next to the $\hbar$ in the path-integral exponent
\begin{equation}
   \bra{\bm{q}_{\rm f}} e^{-\hat{H}\tau/\hbar} \ket{\bm{q}_{\rm i}} = \int^{\bm{q}(\tau)=\bm{q}_{\rm f}}_{\bm{q}(0)=\bm{q}_{\rm i}} \mathcal{D}\bm{q} \, e^{-S_\tau[\bm{q}(\tau')]/(\varepsilon\hbar)}
\end{equation}
and considering the behaviour as $\varepsilon\to0$.
 This prescription is equivalent to a WKB analysis in one dimension. By analogy with Eq.~(\ref{eq:steepest_descent}) one sees that the resulting asymptotic expansion of the path integral will be around a classical trajectory (where the action is stationary).   

There exist two qualitatively different approaches to the derivation of instanton theory: one based on reactive scattering theory,\cite{Miller1975semiclassical,Chapman1975rates,Miller1983rate,Richardson2016FirstPrinciples,Richardson2018InstReview} and the other on the concept of imaginary free-energy (the ``$\Im F$ premise'').\cite{Langer1967ImF,Langer1969ImF,Coleman1977ImF,Callan1977ImF,Coleman1979UsesOfInstantons,Stone1977ImF} 
The $\Im F$ premise was first proposed by Langer in the context of droplet formation,\cite{Langer1967ImF,Langer1969ImF} and then later by Coleman in the context of quantum field theory.\cite{Coleman1977ImF}
Proposed heuristically, the $\Im F$ principle says that the rate of decay of a metastable state is related to the imaginary part of its free-energy as $k= -\frac{2}{\hbar} \Im F$. This is then evaluated using analytic continuation to define the integral over the unstable mode in an asymptotic ($\hbar\to0$) evaluation of the path-integral expression for the partition function. Separately, Miller arrived at instanton theory by starting from the flux-correlation formulation for the thermal rate, using arguments based on Weyl correspondence in combination with semiclassical results from Gutzwiller.\cite{Miller1975semiclassical}
Although the prefactor in Miller's theory appears qualitatively different to that of the $\Im F$ formulation, the two theories have been proven to be equivalent.\cite{Althorpe2011ImF}$^{,}$\footnote{It has been argued recently that there should be an additional term that does not appear in the standard expressions for the instanton rate considered in these works, and that this additional term arises due to non-separability [Y. Georgievskii and S. J. Klippenstein, J. Chem.~Theory Comput.~17, 3863-3885 (2021)]. However, these arguments were not based on a rigorous semiclassical ($\hbar\to0$) analysis. Differences to the standard expressions may, therefore, simply be explained as arising from subdominant terms.} 
More recently Richardson has derived instanton theory from first principles by evaluating the exact flux-correlation expression for the quantum rate asymptotically as $\hbar\to0$.\cite{Richardson2016FirstPrinciples,Richardson2018InstReview} Unlike the $\Im F$ derivation, which is rather subtle and thus difficult to generalise or extend, the flux-correlation formalism provides a rigorous framework for extending the theory, as has been exploited in recent years in the study of electronically nonadiabatic chemical reactions.\cite{inverted,PhilTransA,thiophosgene,nitrene,4thorder,Fang2023ConicalIntersections,Ansari2024Oxygen,Richardson2024NonAdTunneling} For this reason this is the approach taken in the present study.

Despite its success in describing deep tunneling processes, instanton theory has a major issue: it breaks down at high temperature. 
This can  be understood by noting that the shortest period for an instanton orbit is determined by the barrier frequency, $\tau_{\rm c}=2\pi/\omega$.
Hence, Eq.~(\ref{eq:Standard_Instanton}) cannot be applied for temperatures above the same ``crossover temperature'', $T_{\rm c}=\hbar/(k_{\rm B} \tau_{\rm c})$, that appears when considering the parabolic barrier rate. 
This is not a coincidence, as we will see later, it is because the parabolic barrier rate is closely related to the instanton result, as it is the first term in the asymptotic expansion of the rate in powers of $\hbar$ when $T>T_{\rm c}$. 
As one approaches the crossover temperature from below, instanton theory becomes less accurate (although does not diverge).
This is clearly undesirable, as we would like to be able to accurately describe the onset of tunneling in chemical reactions.
 The goal of the present paper is therefore to derive a rigorous uniform semiclassical theory that is valid for all values of the thermal time $\tau=\beta\hbar>0$. Here, \emph{uniform} is a technical term that refers to an expression that is valid for a range of values of an additional (non-asymptotic) parameter that spans two (or more) regions exhibiting qualitatively different asymptotic behaviour. 

That instanton theory breaks down at the crossover temperature was, of course, known by its initial proponents.\cite{Miller1975semiclassical,Coleman1977ImF} As such, there have been many previous attempts to both analyse and ameliorate this problem.\cite{Affleck1981ImF,Grabert1984Crossover,Hanggi1986ImF,Cao1996QTST,Kryvohuz2011rate,McConnell2017instanton,Upadhyayula2023UniformInstanton} In particular, %
in 1981 Affleck argued based on WKB that the $\Im F$ principle should satisfy: $k=-\frac{2}{\hbar} \Im F$ for $T\leq T_c$ and $k=-\frac{\omega\tau}{\pi\hbar} \Im F$ for $T\geq T_c$.\cite{Affleck1981ImF}
These ideas have been developed further by several different authors\cite{Grabert1984Crossover,Hanggi1986ImF,Cao1996QTST,Kryvohuz2011rate,McConnell2017instanton} resulting in a piecewise theory.\cite{Cao1996QTST,Kryvohuz2011rate,McConnell2017instanton}
However, there are a number of drawbacks to this approach.
Being based on a heuristic combination of the $\Im F$ principle and WKB it is difficult to see how to rigorously extend the theory, for example by incorporating higher order asymptotic (i.e.~perturbative)  corrections.\cite{Lawrence2023RPI+PC}
Furthermore, the derivation is based on the fundamental assumption that the instanton smoothly collapses to the transition state with increasing temperature, and hence cannot be generalised to more complex systems.
An alternative approach that has been suggested is to start from the uniform WKB approximation to the 1D microcanonical transmission probability,\cite{Kemble1935WKB,Froman1965JWKB} and then calculate the thermal rate by computing the integral over energy numerically.\cite{Faraday,McConnell2017microcanonical,JoeFaraday} However, these approaches suffer from similar issues to those based on $\Im F$, and require ad-hoc approximations to treat non-separable multidimensional systems.
Recently Upadhyayula and Pollak have proposed a theory that they call ``uniform semiclassical instanton theory'' that replaces this numerical integration with an analytical approximation, which does not exhibit a divergence at the crossover temperature. However, it should be emphasized that the name of the method refers to the use of the uniform WKB transmission probability,\cite{Kemble1935WKB,Froman1965JWKB} and the resulting theory is not a uniform (in $\tau$) asymptotic expansion of the rate as $\hbar\to 0$. This is demonstrated numerically in the supplementary material where it is shown that the theory becomes less accurate near $\tau=\tau_c$ as $\hbar\to0$.

\section{Theory}\label{sec:theory}

In order to derive our general semiclassical rate theory and solve the crossover problem we return to a first-principles derivation of instanton theory. The basis for our approach is the derivation of Richardson from Ref.~\citenum{Richardson2018InstReview}, but with a small change to the flux operators. After a preliminary recap of the quantum flux-flux formulation of rate theory, Sec.~\ref{sec:asymptotic_flux_flux} introduces key definitions and derives the asymptotic form of the flux-flux correlation function. Section~\ref{sec:real_time_instanton_derivation} then discusses how, below the crossover temperature, the change made to the flux operators results in a new real-time perspective on instanton theory. Using this new perspective, Sec.~\ref{sec:crossover_problem} provides a unified understanding of the breakdown of both instanton theory and the parabolic barrier approximation at the crossover temperature. Section~\ref{sec:uniform_asymptotics} then discusses how this unified understanding can be combined with modern methods for deriving uniform asymptotics expansions. 
Completing the derivation, Sec.~\ref{sec:final_theory} introduces the key result of the paper and discusses its behaviour.

The starting point for our derivation is the exact expression for the rate in terms of the time integral of a flux-flux correlation function\cite{Yamamoto1960rate,Miller1974QTST,Chandler1978TST,Miller1983rate}
 \begin{equation}
    k Z_r = \int_0^\infty  c_{\rm ff}(t) \mathrm{d}t, \label{eq:flux-flux_correlation_function}
\end{equation}
where we make use of the most general form of the correlation function, in which the two flux operators can be chosen to be different\cite{Miller1983rate,Miller2003QI}
 \begin{equation}
      c_{\rm ff}(t) =  \tr[e^{-(\tau/2 - it)\hat{H}/\hbar}\hat{F}_p e^{-(\tau/2 + it)\hat{H}/\hbar}\hat{F}_r]. \label{eq:cff_1}
\end{equation}
The flux operators are formally defined as Heisenberg time derivatives of projection operators onto reactants ($\hat{P}_r$) and products ($\hat{P}_p$) as
\begin{subequations}
\begin{align}
   \hat{F}_r &= \frac{-i}{\hbar} \left[\hat{H},\hat{P}_r\right] \\
   \hat{F}_p &= \frac{i}{\hbar} \left[\hat{H},\hat{P}_p\right]
   \end{align}
\end{subequations}
where the projection operators are given by
\begin{subequations}
\begin{align}
   \hat{P}_r &= 1-h(\sigma(\hat{\bm{q}})-s_r) \\
   \hat{P}_p &= h(\sigma(\hat{\bm{q}})-s_p) 
\end{align}
\end{subequations}
here $h(x)$ is the Heaviside step function, such that $\sigma(\bm{q})<s_r$ defines the reactants and $\sigma(\bm{q})>s_p$ the products.
Thus $\hat{F}_r$ measures the flux out of the reactant states (through the ``dividing surface'' $\sigma(\bm{q})=s_r$) and $\hat{F}_p$ measures the flux into the product states (through the ``dividing surface'' $\sigma(\bm{q})=s_p$).

In Richardson's derivation\cite{Richardson2016FirstPrinciples,Richardson2018InstReview} he took both flux operators to be the same and defined the dividing surface to pass through the saddle point of the potential.
Such a choice ensures that the flux correlation function has its maximum at $\Re t=0$.
This is a natural choice when considering instanton theory as an extension of transition state theory, as integrating over time asymptotically (as $\hbar\to0$) then straightforwardly leads to a theory that involves no real-time quantities.
However, in the present work we choose instead to place the dividing surfaces well away from the barrier (out in the reactant and product asymptotes respectively) such that the correlation function reaches its maximum for finite real time. %
The reader might be worried that the presence of real time will mean that we must contend with the infamous sign problem.
This concern would be realised if we tried to evaluate the resulting path-integral expressions exactly or via a quantum instanton approximation\cite{Miller2003QI,Yamamoto2004QI,Ceotto2005QI,Vanicek2005QI,Wang2011QI,QInst,Wolynes1987nonadiabatic,Lawrence2018Wolynes,Lawrence2020NQI} (an idea discussed in Ref.~\citenum{Aieta2017RealTimeQuantumInstanton}).
However, as we will evaluate all integrals analytically using asymptotic analysis our semiclassical theory will have no such issue.
In fact, instead of making the problem more difficult, the presence of real time will actually be the key to solving the problem at the crossover temperature.

\subsection{Asymptotic expression for the flux-flux correlation function}\label{sec:asymptotic_flux_flux}
Before we evaluate the integral over time, we begin by evaluating the flux-flux correlation function asymptotically. The steps we follow in this subsection follow closely those of Ref.~\citenum{Richardson2018InstReview}, and are included here for completeness. However, for the sake of notational simplicity we restrict the discussion here to a system in one dimension.
To further simplify notation we define the forward and backward times, $t_\pm =-i\tau_\pm =\pm t-i\tau/2$, and introduce the imaginary time propagator $\hat{K}(\theta)=e^{-\theta \hat{H}/\hbar}$ such that 
\begin{equation}
   c_{\rm ff}(t) = \tr[\hat{K}(\tau_-)\hat{F}_p \hat{K}(\tau_+) \hat{F}_r]. \label{eq:cff_2}
\end{equation}
Note that $\tau_\pm$ are therefore complex numbers, the imaginary part of which is ($\pm$) the ``real time''.
In one dimension the flux operators can be written in the form
\begin{equation}
    \hat{F}_\alpha = \frac{\hat{p}}{2m}\delta(\hat{q}-s_\alpha)+\delta(\hat{q}-s_\alpha)\frac{\hat{p}}{2m}
\end{equation}
for $\alpha=r$ or $p$, where $\delta(x)$ is the Dirac delta function.
By first inserting resolutions of the identity $1=\int \mathrm{d}q \ketbra{q}{q}$ and then making use of the relation $\bra{q}\hat{p}=-i\hbar \frac{\partial}{\partial q}\bra{q}$ followed by integration over the resulting Dirac delta functions, it is straightforward to show that 
\begin{equation}
\begin{aligned}
    c_{\rm ff}(t) =  \left(\frac{-i\hbar}{2m}\right)^2  \Bigg( &\frac{\partial K(\tau_-,s_p,s_r)}{\partial s_r } \frac{\partial K(\tau_+,s_r,s_p)}{ \partial s_p} \\
     - &\frac{\partial^2 K(\tau_-,s_p,s_r)}{\partial s_r \partial s_p} K(\tau_+,s_r,s_p) \\
     - &K(\tau_-,s_p,s_r)\frac{\partial^2 K(\tau_+,s_r,s_p)}{ \partial s_r \partial s_p}\\
     +&\frac{\partial K(\tau_-,s_p,s_r)}{\partial s_p } \frac{\partial K(\tau_+,s_r,s_p)}{ \partial s_r} \Bigg), \label{eq:exact_cff_1D_integrated}
\end{aligned}
\end{equation}
where the position representation of the propagator is defined as
\begin{equation}
    \bra{q''}\hat{K}(\theta)\ket{q'}={K}(\theta,q',q'') . \label{eq:propagator_definition}
\end{equation}
Now we turn to the asymptotic evaluation of this exact expression as $\hbar\to0$. This can be done numerically by first writing the propagator as a discretised path integral and then evaluating the integrals over position by steepest descent [i.e.~using Eq.~(\ref{eq:steepest_descent})]. For the present purpose, we note that in the continuum limit this is formally equivalent to using the semiclassical (imaginary-time) van-Vleck propagator       
\begin{equation}
    \bra{q''}\hat{K}(\theta)\ket{q'}={K}(\theta,q',q'') \sim \sum_{\rm paths} \left(\frac{C}{2\pi\hbar}\right)^{\nicefrac{1}{2}} e^{-S(\theta,q',q'')/\hbar} \label{eq:van-Vleck-1D}
\end{equation}
where $S(\theta,q',q'')$ is the Euclidean action calculated along the classical trajectory from $q'$ to $q''$ in total imaginary time $\theta$, \begin{equation}
    S(\theta,q',q'') = \int_0^\theta  \frac{1}{2}m \dot{q}^2(\theta')+V\!\left(q(\theta')\right) \, \mathrm{d}\theta', \label{eq:action_definition_1D}
\end{equation}
the prefactor is defined as
\begin{equation}
    C =  -\frac{\partial^2 S}{\partial q' \partial q''},
\end{equation}
and the sum is over all classical paths that go from $q'$ to $q''$ in time $\theta$. Note the branch of the square root in the prefactor is chosen so that the prefactor is continuous along the trajectory.\footnote{The prefactor is of course a property of the entire trajectory. The prefactor being continuous along the trajectory should be understood as corresponding to considering the set of prefactors that are given by fixing $q'$ and then varying $\theta$ and $q''$ such that they follow the trajectory, and ensuring that this set is continuous for some parameterisation of the trajectory.}
In addition to the asymptotic expression for the propagator we will also need to make use of the asymptotic expressions for its derivatives\footnote{Note that terms involving derivatives of the prefactor are subdominant and hence are not included.}
\begin{subequations}
\begin{align}
    \frac{\partial{K}}{\partial q'} &\sim \sum_{\rm paths} -\frac{1}{\hbar} \frac{\partial S}{\partial q'} \left(\frac{C}{2\pi\hbar}\right)^{\nicefrac{1}{2}} e^{-S(\theta,q',q'')/\hbar}  \label{eq:van-Vleck-1D_derivative_a} \\
  \frac{\partial{K}}{\partial q''} &\sim  \sum_{\rm paths}-\frac{1}{\hbar} \frac{\partial S}{\partial q''} \left(\frac{C}{2\pi\hbar}\right)^{\nicefrac{1}{2}} e^{-S(\theta,q',q'')/\hbar} \label{eq:van-Vleck-1D_derivative_b} \\
    \frac{\partial^2{K}}{\partial q' \partial q''} &\sim \sum_{\rm paths} \frac{1}{\hbar^2} \frac{\partial S}{\partial q'}\frac{\partial S}{\partial q''} \left(\frac{C}{2\pi\hbar}\right)^{\nicefrac{1}{2}} e^{-S(\theta,q',q'')/\hbar}. \label{eq:van-Vleck-1D_derivative_c}
\end{align}
\end{subequations}
Combining Eq.~(\ref{eq:exact_cff_1D_integrated}) with Eqs.~(\ref{eq:van-Vleck-1D})-(\ref{eq:van-Vleck-1D_derivative_c}) and retaining just the dominant path we then obtain the following asymptotic expression for the correlation function (valid as $\hbar\to0$)
\begin{equation}
\begin{aligned}
    c_{\rm ff}(t) \sim  &\Bigg(\frac{\partial S_{\! -}}{\partial s_r}\frac{\partial S_{\! +}}{\partial s_p}-\frac{\partial S_{\! -}}{\partial s_p}\frac{\partial S_{\! -}}{\partial s_r}
    -\frac{\partial S_{\! +}}{\partial s_p}\frac{\partial S_{\! +}}{\partial s_r}+\frac{\partial S_{\! -}}{\partial s_p}\frac{\partial S_{\! +}}{\partial s_r}\Bigg)\\ &\times\frac{-1}{4m^2} \sqrt{\frac{C_+C_-}{(2\pi\hbar)^2}} e^{-[S_{\! +}(\tau_+,s_r,s_p)+S_{\! -}(\tau_-,s_p,s_r)]/\hbar},
    \end{aligned}
\end{equation}
where we have labelled the action (and associated quantities) $\pm$ to distinguish between the forward and backward paths. Hence, defining the total Euclidean action as the sum of the forward and backward parts (suppressing the dependence on $\tau$ and $s_\alpha$)
\begin{equation}
    S(t) = S_{\! +}(\tau/2+it,s_r,s_p) + S_{\! -}(\tau/2-it,s_p,s_r) \label{eq:S_t_definition}
\end{equation}
and the $\hbar$ independent part of the prefactor as
\begin{equation}
    A(t) \!=\! \frac{(C_+C_-)^{\nicefrac{1}{2}}}{8m^2\pi}\Bigg(\frac{\partial S_{\! -}}{\partial s_p}\frac{\partial S_{\! -}}{\partial s_r}
    -\frac{\partial S_{\! -}}{\partial s_r}\frac{\partial S_{\! +}}{\partial s_p}-\frac{\partial S_{\! -}}{\partial s_p}\frac{\partial S_{\! +}}{\partial s_r}+\frac{\partial S_{\! +}}{\partial s_p}\frac{\partial S_{\! +}}{\partial s_r}\Bigg)
\end{equation}
we see that the correlation function has the simple asymptotic form
\begin{equation}
    c_{\rm ff}(t) \sim \frac{A(t)}{\hbar} e^{-S(t)/\hbar}. \label{eq:asymptotic_flux_flux_form}
\end{equation} 
Note that, while the preceding analysis was specific to a one-dimensional system, 
the flux-flux correlation function for multidimensional systems also has the same form as Eq.~(\ref{eq:asymptotic_flux_flux_form}).

\subsection{A real-time derivation of instanton theory}\label{sec:real_time_instanton_derivation}
Combining Eq.~(\ref{eq:flux-flux_correlation_function}) with Eq.~(\ref{eq:asymptotic_flux_flux_form}) we can recover the standard instanton theory below the crossover temperature by integrating over time asymptotically using Eq.~(\ref{eq:steepest_descent}) to obtain
\begin{equation}
    k Z_r \sim \frac{A(t^\star)}{\hbar} \sqrt{\frac{2\pi\hbar}{S''(t^\star)}}e^{-S(t^\star)/\hbar} \text{ as } \hbar\to 0, \label{eq:Standard_Instanton_RT}
\end{equation}
where $t^\star$ satisfies the steepest descent condition $S'(t^\star)=0$. We will now argue that, $S(t^\star)=S_{\! \rm inst}(\tau)$, and that Eq.~(\ref{eq:Standard_Instanton_RT}) is equivalent to the standard instanton theory. 

First note that $S'(t^\star)=0$ implies 
\begin{equation}
    \frac{\partial S_{\! +}}{\partial \tau_+} - \frac{\partial S_{\! -}}{\partial \tau_-} = E_+ - E_- =0 \label{eq:stationary_action_condition}
\end{equation}
i.e.~the energy of the forward and backward paths, $E_{\pm}$, must be the same. Further, since $K(\theta,q',q'')=K^*(\theta^*,q'',q')$ it follows that, on the real $t$ axis, the action for the forward and backward paths are always complex conjugates $S_{\! +}(\tau_+,s_r,s_p)=S^*_{\!-}(\tau_-,s_p,s_r)$ making $S(t)$ real. Differentiating this with respect to $\tau$ then gives $E_+=E_-^*$ which combined with Eq.~(\ref{eq:stationary_action_condition}) shows that the energy of the forward and backward trajectories at the stationary time are also real.

Before we discuss finding a trajectory that satisfies these conditions, note that there is a freedom we have not yet discussed: the contour of integration (the time path) in the definition of the action [Eq.~(\ref{eq:action_definition_1D})]. This freedom corresponds to the order of the real- and imaginary-time propagators in the path-integral discretisation of Eq.~(\ref{eq:propagator_definition}). Of course the exact expression is independent of this choice because real- and imaginary-time propagators commute. The semiclassical propagator must necessarily retain this property. This can be seen explicitly in Eq.~(\ref{eq:action_definition_1D}) as being a result of Cauchy's integral theorem. We are therefore free to choose the most convenient time path for our purposes (under the assumptions of the theorem). 

\begin{figure}[t]
    \centering
    \includegraphics[width=1.0\linewidth]{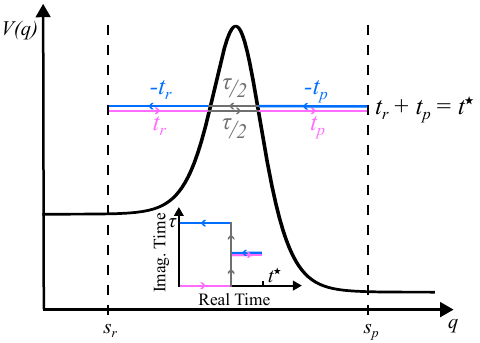}
    \caption{Diagram depicting the dominant trajectory below the crossover temperature, $\tau>\tau_c$. Starting at the reactants, the system first moves in real time for $t_r$ up to the point where its momentum is zero (the turning point). It then ``tunnels'' through the barrier by moving in imaginary time for $\tau/2$ along the instanton trajectory to the other turning point. The trajectory then carries on to the product dividing surface in real time for $t_p$. Finally, the trajectory returns from the product dividing surface to the reactant dividing surface following the same path but with each of the real time segments now corresponding to negative real time.}
    \label{fig:trajectory_diagram}
\end{figure}

With a careful choice of time path it is 
 trivial to find a trajectory that satisfies $S(t^\star)=S_{\! \rm inst}(\tau)$\@. Figure~\ref{fig:trajectory_diagram} depicts the corresponding trajectory and time path. 
 The forward and backward time paths each consist of three segments; two pure real-time segments either side of a pure imaginary-time segment of length $\tau/2$. 
Rather than defining the lengths of the real time segments ($t_r$ and $t_p$) we instead define the trajectory along the imaginary-time segment to be half of the instanton orbit from one turning point to the other.
This then uniquely determines the real-time sections of the trajectories along with the times $t_r$ and $t_p$. To see this note that, because the momentum is continuous along the trajectory it must be zero at the points connecting the real-time and imaginary-time segments. Hence, $t_r$ ($t_p$) must be the time it takes for the system to roll from the reactant (product) end of the instanton to the reactant (product) dividing surface.  This, therefore, uniquely determines the stationary time as $t^\star=t_r+t_p$. Since the forward and backward trajectories are entirely real and follow the same path it is clear that $E_+=E_-$ and hence [cf.~Eq.~(\ref{eq:stationary_action_condition})] the trajectory corresponds to a stationary time, $t^\star$.

The stationary trajectory gives an intuitive picture of 
the reaction process. 
The system starts at the reactants and moves in real time towards the barrier. Upon reaching the turning point of the trajectory at the barrier, instead of bouncing off the barrier in real time it switches to imaginary time. This effectively ``turns the barrier upside-down'' allowing the system to tunnel through to the product side. The system then switches back to real time and carries on to the products. Because of the cyclic nature of the trace in Eq.~(\ref{eq:cff_1}) the system then retraces its steps, moving backwards in real time to the barrier before tunneling through the barrier in (positive) imaginary time and then back to the reactants again in negative real time. As the forward and backward real-time segments follow the same paths, their contributions to the action exactly cancel one another leaving only the imaginary time contribution, and hence $S(t^\star)=S_{\! \rm inst}(\tau)$. 
Importantly, the uniqueness of asymptotic power series means that the prefactor must also be equivalent to the usual instanton prefactor.  
Hence, we have that 
\begin{equation}
\frac{A(t^\star)}{\hbar} \sqrt{\frac{2\pi\hbar}{S''(t^\star)}} = \frac{Z_{\rm inst}(\tau)}{\sqrt{2\pi\hbar}} \left(-\frac{\mathrm{d}^2S_{\!\mathrm{inst}}}{\mathrm{d}\tau^2}\right)^{\nicefrac{1}{2}}.
\end{equation}

For the inquisitive reader, a short aside: There are two obvious questions based on the preceding discussion. First, what would happen to the trajectory if we kept $t^\star$ (and $\tau$) fixed but deformed the time path? The answer is that the resulting trajectory must move into the complex position plane. This can be understood by noting that, if the momentum of the system is real then propagation in anything other than real time will lead to a change in the position that is complex. As the initial and final momenta are fixed changing the direction of the time path at any point (other than turning points) will, therefore, result in a complex trajectory. The second question is, when $t$ is not $t^\star$ can a time path still be found that keeps the trajectory real? The answer is that while this is possible for some $t$ and $\tau$ in one dimension, it is generally not possible in multiple dimensions. Finally, the observant reader may note that there is another stationary trajectory that swaps the sign of the real time segments on the product side of the barrier to give $t^\star=t_r-t_p$. One might, therefore, wonder why this trajectory doesn't also contribute. A careful consideration, however, shows that this corresponds to a different branch of the correlation function and hence does not need to be included.
\subsection{Diagnosing the problem at the crossover temperature} \label{sec:crossover_problem}
\begin{figure}[t]
    \centering
    \includegraphics[width=1.0\linewidth]{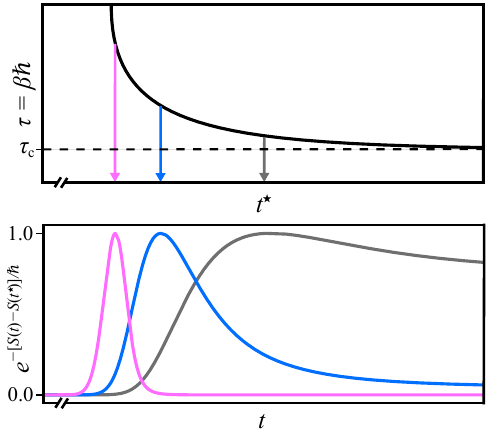}
    \caption{Plot of the steepest descent time, $t^\star$, as a function of the inverse temperature (period of the instanton orbit, $\tau$) for $\tau>\tau_{\rm c}$. }
    \label{fig:approaching_crossover}
\end{figure}

With this new real time perspective we can now obtain a simple intuitive picture of what happens as we approach the crossover temperature, and hence why the standard instanton theory stops working there. First consider approaching the crossover temperature from below.
Figure~\ref{fig:approaching_crossover} shows how the steepest descent time, $t^\star$, varies with $\tau$ for a typical system close to the crossover. We see that as  $\tau\to\tau_{\rm c}$ the stationary time approaches infinity, $t^\star\to\infty$. This can be understood intuitively by noting that as $\tau\to\tau_{\rm c}$ the turning points, where the real-time and imaginary-time parts of the path meet, get closer and closer to the top of the barrier. As this happens, the force at the turning points approaches zero. Hence, the  real-time segments have to become longer and longer to give time for the system roll away from (or come to a stop at) the barrier.

Now we consider the behaviour above the crossover temperature. Based on our preceding discussion it is clear that in this regime the correlation function must be dominated by the behaviour near $t=\infty$. For a one-dimensional barrier we show in the supplementary material (following Ref.~\citenum{ZinnJustin2021BookCh37}) that for large values of real time the action behaves like
\begin{equation}
    S(t) \sim S_{\! \infty} + a \sin(\omega \tau/2)\, e^{-\omega t} \text{ as } t\to\infty \label{eq:action_at_infinity}
\end{equation}
where $S_{\! \infty}=\tau V^\ddagger$ and $a$ is a constant with units of action. Furthermore, the pre-exponential factor varies as 
\begin{equation}
    A(t) \sim \frac{a\omega^2}{4\pi} e^{-\omega t} \text{ as } t\to\infty. \label{eq:prefactor_at_infinity}
\end{equation}
Hence, the correlation function obeys
\begin{equation}
    c_{\rm ff}(t) \sim  \frac{a\omega^2}{4\pi \hbar} e^{-\omega t} \exp(-\left[S_{\! \infty} + a \sin(\frac{\omega \tau}{2})\, e^{-\omega t}\right]/\hbar) \label{eq:cff_asymptotic_above_crossover}
\end{equation}
as $\hbar\to0$ and $t\to\infty$.
Upon making the substitution $v=e^{-\omega t}$ we can see that (above the crossover temperature) the rate is asymptotic to\footnote{Note that after making the substitution we also extend the integration range from $0$ to $1$ to be from $0$ to $\infty$. This is the correct thing to do because as $\hbar\to0$ the integrand becomes more and more narrowly peaked at the boundary $v=0$.}
\begin{equation}
\begin{aligned}
    k Z_r &\sim \int_0^\infty \frac{a\omega}{4\pi \hbar}  \exp(-\left[S_{\! \infty} + a \sin(\frac{\omega \tau}{2})\, v\right]/\hbar) \mathrm{d}v \\ &= \frac{\omega}{4\pi\sin(\omega\tau/2)}e^{-S_{\! \infty}/\hbar} \equiv \frac{\omega}{4\pi\sin(\beta\hbar\omega/2)} e^{-\beta V^\ddagger} \label{eq:Parabolic_Barrier}
\end{aligned}
\end{equation}
which is the well-known parabolic barrier rate.\cite{BellBook} Note that this expression differs qualitatively in its dependence on (the asymptotic parameter associated with) $\hbar$ from the instanton result evaluated below the crossover temperature, Eq.~(\ref{eq:Standard_Instanton_RT}). 

We can now gain a unified perspective on the behaviour both above and below crossover. To do so we begin by making the same variable transformation as above, $v=e^{-\omega t}$, such that the rate can be expressed as
\begin{equation}
    k Z_r \sim \int_0^1 \frac{G(v)}{ \hbar} e^{-S(v)/\hbar} \mathrm{d} v  \text{ as } \hbar\to0
\end{equation}
where we define $G(v):=A(t(v))/(v\omega)$ and $S(v):=S(t(v))$.\footnote{Importantly, with this definition $G(v)$ is a smooth function of $v$ for all $v\geq0$.} It is important to recognise that we will still recover the standard instanton result if we perform the integral over $v$ asymptotically below the crossover temperature [i.e.~using Eq.~(\ref{eq:steepest_descent})]. 
This can be seen explicitly by first integrating asymptotically over $v$ to give
\begin{equation}
    k Z_r \sim \frac{G(v^\star)}{\hbar} \sqrt{\frac{2\pi\hbar}{S''(v^\star)}}e^{-S(v^\star)/\hbar} \text{ as } \hbar\to 0
\end{equation}
and then using the chain rule  $-v\omega S'(v)=S'(t)$ to show that $S(t^\star)=S(v^\star)$ and $(v^\star\omega)^2 S''(v^\star)=S''(t^\star)$.
The advantage of making this transformation is that the behaviour at the crossover temperature becomes very simple. It just corresponds to the temperature at which the stationary point, $v^\star$, moves from being inside to outside of the integration range, as we will now illustrate.   

\begin{figure}[t]
    \centering
    \includegraphics[width=1.0\linewidth]{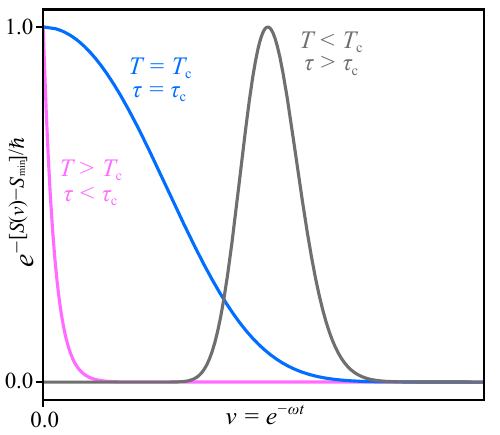}
    \caption{Plot of $e^{-S(v)/\hbar}$, normalised by its maximum value in the integration range, showing an example for $\tau<\tau_c$, $\tau=\tau_c$ and $\tau>\tau_c$.}
    \label{fig:uniform_diagram1}
\end{figure}

Figure \ref{fig:uniform_diagram1} depicts $e^{-S(v)/\hbar}$ for three temperatures, one above crossover, one below crossover, and the other exactly at crossover. We see that, well below the crossover temperature, $e^{-S(v)/\hbar}$ is peaked far away from the boundary, and hence approximating the integrand as a Gaussian is a reasonable approximation. Equally, well above the crossover temperature we see that only the tail of the function $e^{-S(v)/\hbar}$ appears inside the integration region, and hence approximating the integrand with an exponential [as is done in Eq.~(\ref{eq:Parabolic_Barrier})] is also a reasonable approximation. However, exactly at the crossover temperature the stationary point $S'(v^\star)=0$ occurs exactly on the integration boundary, $v^\star=0$. 
This means that essentially half of the peak is outside of the integration range.
Hence, instanton theory (which assumes the full peak is inside the integration bounds) is approximately a factor of two too large. When approaching the crossover temperature from above things are much worse, as the first derivative at the boundary [i.e.~the coefficient of the term linear in $v$ in the exponent of Eq.~(\ref{eq:Parabolic_Barrier})] approaches zero and hence an exponential approximation will give a divergent integral.       

\subsection{Uniform asymptotics} \label{sec:uniform_asymptotics}
To obtain a rate theory that is valid both above and below the crossover temperature we need to make use of ideas from uniform asymptotics. A uniform asymptotic expression is one that is continuous in a parameter (distinct from the asymptotic parameter) that connects two or more regions of different asymptotic behaviour. We have seen in the previous section that, as $\tau$ varies the stationary point moves from inside to outside of the integration range, resulting in a qualitative change in the form of the asymptotic approximation for the rate. We are, therefore, interested in finding an asymptotic expression for the rate that is uniform in $\tau$.

A common approach is to construct uniform asymptotic approximations in a piecewise manner. Following this approach one might be tempted to suggest that, below the crossover temperature,  the integral over $v$ should be approximated by the Gaussian integral 
\begin{equation}
    k Z_r \sim \frac{G(v^\star)}{\hbar} \int_0^\infty e^{-[S(v^\star)+(v-v^\star)^2S''(v^\star)/2]/\hbar} \mathrm{d}v, \label{eq:Bad_uniform_1}
\end{equation} 
and that this should then be combined with an expression valid above crossover that has the same value as $\tau\to\tau_{\rm c}$.
This kind of uniform approximation is employed for example in Ref.~\citenum{Kryvohuz2011rate}.
The resulting expression has lots of the behaviour that one expects, it reduces the prediction by a factor of $\nicefrac{1}{2}$ at the crossover temperature and smoothly approaches the standard result as one lowers the temperature for fixed $\hbar$ (and also as one lowers $\hbar$ for fixed $\tau$). 
However, this is not the correct approach. 

To see what is wrong with Eq.~(\ref{eq:Bad_uniform_1}), we can evaluate the integral and make use of the chain rules given earlier to show that it is equivalent to 
\begin{equation}
\begin{aligned}
    kZ_r \sim \frac{A(t^\star)}{ \hbar} \sqrt{\frac{2\pi\hbar}{S''(t^\star) }} e^{-S(t^\star)/
    \hbar} \frac{\mathrm{erfc}\!\left(- \sqrt{\frac{S''(t^\star) }{2\omega ^2\hbar}} \right)}{2}
    \end{aligned}
\end{equation}
where $\mathrm{erfc}(x)$ is the complementary error function. Crucially, we see that the difference between this and the standard instanton result is the factor,
\begin{equation*}
    \frac{1}{2}\mathrm{erfc}\!\left(- \sqrt{\frac{S''(t^\star) }{2\omega ^2\hbar}} \right).
\end{equation*}
While this factor takes on values between $\nicefrac{1}{2}$ and $1$, the precise value is dependent on the ratio $\frac{S''(t^\star) }{2\omega ^2\hbar}$. This is a problem because the actual value of $S''(t^\star)$ is determined by the location of the dividing surfaces. This is clearly unphysical as the true quantum rate is independent of the choice of dividing surface. [As mentioned earlier, the standard instanton result is also independent of dividing surface. Furthermore, since they are determined by the properties of the saddle point, Eyring-TST and the parabolic barrier rate are also dividing surface independent.] Clearly we want our uniform asymptotic result to also have this property.

Fortunately, if one consults a textbook on uniform asymptotics\cite{RWong1989UniformAsymptotics} one will see that Eq.~(\ref{eq:Bad_uniform_1}) is not the recommended approach. 
Instead, 
one should use Bleistein's method\cite{Bleistein1966Uniform} which has become the standard method for generating uniform asymptotic series.\cite{RWong1989UniformAsymptotics} Our final result will then be the first term in this series, just as the standard instanton result is 
the first term in a regular asymptotic series. This has the advantage that the theory will not only be continuous in $\tau$, but it will also be smooth and continuously differentiable.

Following Bleistein's method we begin by defining a new variable transformation
\begin{equation}
    S(v) = \frac{u^2}{2} - bu + c
\end{equation}
and choosing the constants $b$ and $c$ so that $u(v=0)=0$ and $u(v^\star)=b$. This then implies that
\begin{subequations}
\begin{equation}
    b =  \mathrm{sgn}(v^\star)\sqrt{2S_{\! \infty}-2S(v^\star)} %
\end{equation}
\begin{equation}
    c = S(v\!=\!0) = S_{\! \infty}.
\end{equation}
\end{subequations}
Note that we assume here that $S(v)$ has been analytically continued to $v<0$ such that for $\tau<\tau_{\rm c}$ we can find $v^\star<0$.
Solving for $u$ we then obtain
\begin{equation}
    u = b +\mathrm{sgn}(v-v^\star) \sqrt{2S(v)-2S(v^\star)}
\end{equation}
where the $\mathrm{sgn}(v-v^\star)$ ensures that the variable transform is single valued. After this variable transformation we have
\begin{equation}
    k Z_r \sim e^{-c/\hbar}\int_0^\infty \frac{g(u)}{ \hbar} e^{-(u^2/2-bu)/\hbar}  \mathrm{d} u \,  \text{ as }\, \hbar\to0
\end{equation}
where $g(u) = G\big(v(u)\big) \frac{\mathrm{d}v}{\mathrm{d}u}$.
The final step in the Bleistein method is to write this pre-exponential term as a linear function that passes through the function at the boundary and the stationary point ($u=b$) plus a remainder term 
\begin{equation}
    g(u) = g(b) + (u-b)\frac{g(b)-g(0)}{b} + r(u).
\end{equation}
As the remainder $r(u)$ is zero at both the stationary point and the boundary it can be ignored at leading order in the uniform asymptotic expansion. With this we can then perform all integrals analytically to obtain
\begin{equation}
\begin{aligned}
    kZ_r &\sim e^{-c/\hbar}\frac{g(b)}{\hbar} \sqrt{2\pi\hbar}\,e^{\phantom{|}b^2/2\hbar}\frac{1}{2}\mathrm{erfc}\!\left(\frac{-b}{\sqrt{2\hbar}}\right)\\
&+e^{-c/\hbar}\frac{g(b)-g(0)}{b} .
\end{aligned}
\end{equation}

To analyse this result we need to express it explicitly in terms of $S(t)$ and $A(t)$.
We begin by defining $S_{\! \infty}-S(t^\star)=\Delta S$ and noting that $\mathrm{sgn}(v^\star)$ can be rewritten in terms of $\tau$ to give
\begin{equation}
    b =\mathrm{sgn}\!\left(\tau-\tau_{\rm c}\right)\sqrt{2\Delta S}.
\end{equation}
Second, we note that $b^2/2$ can be combined with $c$ to give the instanton action
\begin{equation}
    -b^2/2 + c = S(t^\star) = S_{\!\rm inst}(\tau).
\end{equation}
To complete our simplifications we need to determine explicit expressions for $g(b)$ and $g(0)$. This can be achieved by observing that 
\begin{equation}
    \frac{\mathrm{d}u}{\mathrm{d}v} = \mathrm{sgn}(v-v^\star)\frac{S'(v)}{\sqrt{2 S(v)-2S(v^\star)}}
\end{equation}
which can then be combined with the definition of $g(u)$ to give
\begin{equation}
\begin{aligned}
    g(u) &= G(v)\frac{\mathrm{d}v}{\mathrm{d}u} = -A(t)\frac{\mathrm{d}t}{\mathrm{d}v}\frac{\mathrm{d}v}{\mathrm{d}u}\\
    &=\mathrm{sgn}(v^\star-v)A(t) \frac{\sqrt{2S(t)-2S(t^\star)}}{S'(t)}.
    \end{aligned}
\end{equation}
Hence, evaluating this at $u=0$ and taking the limit as $u\to b$ from above or below gives
\begin{subequations}
\begin{equation}
\begin{aligned}
    g(0)&= \mathrm{sgn}\!\left(\tau-\tau_{\rm c}\right) \sqrt{2\Delta S} \lim_{t\to\infty} \frac{A(t)}{S'(t)} %
\end{aligned}
\end{equation}
\text{and}
\begin{equation}
    g(b) = \frac{A(t^\star)}{\sqrt{S''(t^\star)}}. %
\end{equation}
\end{subequations}
Combining these results we obtain 
\begin{equation}
\begin{aligned}
  \!\!  k Z_r &\sim   e^{-S_{\! \infty}/\hbar}\left(   \mathrm{sgn}\!\left(\tau-\tau_{\rm c}\right) \frac{A(t^\star)}{\sqrt{2 S''(t^\star)\Delta S }}-\lim_{t\to\infty} \frac{A(t)}{S'(t)} \right)\\
    &+e^{-S(t^\star)/\hbar}\frac{A(t^\star)}{\hbar} \sqrt{\frac{2\pi\hbar}{S''(t^\star)}}\, \frac{1}{2}\mathrm{erfc}\!\left(\mathrm{sgn}\!\left(\tau_{\rm c}-\tau\right)\!\sqrt{\frac{\Delta S}{\hbar}}\right)  \! \label{eq:First_Key_Result_1}
\end{aligned}
\end{equation}
which is a rigorous uniform asymptotic expression for the thermal rate constant that correctly bridges between the parabolic barrier result above the crossover temperature and the instanton result below crossover.

Although we motivated the derivation in terms of a one-dimensional system, this expression is valid for any system in which the instanton collapses smoothly to the transition state. Furthermore, while we have used the real-time formulation to derive the theory, it can immediately be rewritten in terms of quantities that involve only imaginary time. Hence, it is clearly independent of the choice of dividing surface.
Making use of the relations
\begin{subequations}
\begin{equation}
\frac{A(t^\star)}{\sqrt{S''(t^\star)}} = \frac{Z_{\rm inst}(\tau)}{2\pi} \left(-\frac{\mathrm{d}^2S_{\!\mathrm{inst}}}{\mathrm{d}\tau^2}\right)^{\nicefrac{1}{2}}
\end{equation}
\text{and} 
\begin{equation}
      -\lim_{t\to\infty} \frac{A(t)}{S'(t)} =  \frac{\omega}{4\pi\sin(\tau\omega/2)}Z^\ddagger(\tau)
\end{equation}
\end{subequations}
allows us to write
\begin{equation}
\begin{aligned}
  \!\!  k Z_r &\sim   e^{-\tau V^\ddagger/\hbar}   \mathrm{sgn}\!\left(\tau-\tau_{\rm c}\right)\frac{Z_{\rm inst}(\tau)}{2\pi\sqrt{2\Delta S}} \left(-\frac{\mathrm{d}^2S_{\!\mathrm{inst}}}{\mathrm{d}\tau^2}\right)^{\nicefrac{1}{2}} \\
&+e^{-\tau V^\ddagger/\hbar} \frac{\omega}{4\pi\sin(\tau\omega/2)}Z^\ddagger(\tau) \\
    &+e^{-S_{\!\mathrm{inst}}(\tau)/\hbar}\frac{Z_{\rm inst}(\tau)}{\sqrt{2\pi\hbar}} \left(\!-\frac{\mathrm{d}^2S_{\!\mathrm{inst}}}{\mathrm{d}\tau^2}\right)^{\!\nicefrac{1}{2}}\! \frac{1}{2}\mathrm{erfc}\!\left(\mathrm{sgn}\!\left(\tau_{\rm c}-\tau\right)\!\sqrt{\frac{\Delta S}{\hbar}}\right)  \! 
\end{aligned}\label{eq:First_Key_Result_2a}
\end{equation}
which is equivalent to the more compact expression
\begin{equation}
\begin{aligned}
    k  &\sim  k_{\text{m-pb}}    +\mathrm{sgn}\!\left(\tau-\tau_{\rm c}\right)\frac{k_{\rm inst}e^{-\Delta S/\hbar}}{\sqrt{4\pi\Delta S/\hbar}} \\
    &+k_{\rm inst}\, \frac{1}{2}\mathrm{erfc}\!\left(\mathrm{sgn}\!\left(\tau_{\rm c}-\tau\right)\!\sqrt{\frac{\Delta S}{\hbar}}\right)\label{eq:First_Key_Result_2}
\end{aligned}
\end{equation}
where $k_{\text{m-pb}}$ is the multidimensional parabolic barrier rate.

We are very nearly done with our derivation, however, the keen-eyed reader will note that Eqs.~(\ref{eq:First_Key_Result_1}), (\ref{eq:First_Key_Result_2a}), and (\ref{eq:First_Key_Result_2}) are still not valid at all temperatures. This can be seen by noting that $(\sin(\tau\omega/2))^{-1}$ in the parabolic barrier rate, $k_{\text{m-pb}}$,  diverges not only at $\tau_{\rm c}=2\pi/\omega$, but at all $\tau_{{\rm c},n}=2\pi n/\omega$ for $n=1,2,\dots,\infty$, i.e.~half the crossover temperature, a third of the crossover temperature, and so on. Again the real-time formulation makes it easy to understand the cause of these divergences. Each one corresponds to a different stationary point of the action passing through the $v=0$ boundary.

We can give a physical interpretation to these stationary points by observing that $\tau_{\mathrm{c},n}$ corresponds to the shortest possible time for a trajectory which completes $n$ orbits on the upturned potential. Hence, we can attribute these stationary points to the trajectories that involve multiple orbits. 
For example, when $\tau_{c,3}>\tau>\tau_{c,2}$ ($6\pi/\omega>\tau>4\pi/\omega$) there exists not one but two periodic trajectories on the upturned potential with a period $\tau$. One is the usual instanton, that orbits just once, and the other is a trajectory that orbits twice in the same time, with each half of the trajectory following the same path as the standard instanton at twice the temperature. 

These multi-orbit instantons (typically referred to as periodic instantons) have a higher action and hence are exponentially suppressed compared to the one-orbit instanton. Within standard (Poincar\'e) asymptotics these terms are, therefore, not included as they are smaller than every term in the one-orbit instanton's asymptotic series (as $\hbar\to0$). However, we now see that in order to obtain a rigorous uniform theory valid at any temperature we are naturally led to include them.

\subsection{Semiclassical instanton theory valid at any temperature}\label{sec:final_theory}

Given the preceding discussion it is clear that in order to cancel the divergences in $k_{\text{m-pb}}$ we must modify Eq.~(\ref{eq:First_Key_Result_2}) by including multi-orbit instanton terms. The form of these terms can easily be determined by inspection. However, we do not have to rely on inspection alone. 
As shown in the Appendix, in one dimension we can rigorously derive the desired uniform thermal rate theory in an entirely different way. There, we start from the uniform energy dependent WKB transmission probability and then obtain the thermal rate by integrating over energy asymptotically using Bleistein's method.\cite{Bleistein1966Uniform} 
The resulting theory contains exactly the multi-orbit terms we were expecting. Comparing this one-dimensional result  [Eq.~(\ref{eq:one-dimensional-result})] with  Eq.~(\ref{eq:First_Key_Result_2}) it is trivial to generalise to the multidimensional multi-orbit case. 
We thus arrive at the central result of the present paper, a uniform asymptotic expression for the thermal rate in multiple dimensions
\begin{equation}
\begin{aligned}
    k  &\sim   k_{\text{m-pb}}+ \sum_{n=1}^\infty (-1)^{n+1} \mathrm{sgn}\!\left(\tau-n\tau_{\rm c}\right) \frac{k_{n,\rm inst}e^{-\Delta S_{\!n}/\hbar}}{\sqrt{4\pi\Delta S_{\!n}/\hbar}}  \\
    &+\sum_{n=1}^\infty (-1)^{n+1} k_{n,\rm inst}\, \frac{1}{2}\mathrm{erfc}\!\left(\mathrm{sgn}\!\left(n\tau_{\rm c}-\tau\right)\!\sqrt{\frac{\Delta S_{\! n}}{\hbar}}\right)  \label{eq:Full_Uniform_Theory_simplified}
\end{aligned}
\end{equation}
where the $n$-orbit instanton action is defined as
\begin{equation}
    S_{\!n,\mathrm{inst}}(\tau) = n S_{\!\rm inst}(\tau/n)
\end{equation}
and the difference to the collapsed/classical action is given by
\begin{equation}
    \Delta S_{\!n} = \tau V^\ddagger - S_{\!n,\mathrm{inst}}(\tau).
\end{equation}
The only remaining terms to define are the effective $n$-orbit instanton rate constants that are given by
\begin{equation}
    k_{n,\mathrm{inst}}Z_r = e^{-S_{\!n,\mathrm{inst}}(\tau)/\hbar}\frac{Z_{n,\mathrm{inst}}(\tau)}{\sqrt{2\pi\hbar}} \left(-\frac{\mathrm{d}^2S_{\!n,\mathrm{inst}}}{\mathrm{d}\tau^2}\right)^{\nicefrac{1}{2}}
\end{equation}
with $Z_{n,\mathrm{inst}}(\tau)$ the instanton partition function for the $n$-orbit instanton. 
In the absence of rotations this is given by
\begin{equation}
    Z_{n,\rm inst}(\tau) = \prod_{j=1}^{f-1} \frac{1}{2\sinh[n u_j(\tau/n)/2]},
\end{equation}
where again $u_j(\tau)$ is the $j^{\rm th}$ stability parameter for the 1-orbit instanton of period $\tau$. %
Note that, just as with the standard instanton theory, Eq.~(\ref{eq:Full_Uniform_Theory_simplified}) is rigorously independent of dividing surface and satisfies detailed balance.
We stress again that, although the theory is applicable to a wide range of multidimensional systems, our analysis, and hence Eq.~(\ref{eq:Full_Uniform_Theory_simplified}), assumes that the instanton collapses smoothly to the transition state as the temperature is increased. Notable exceptions, such as quartic barriers, where there are multiple (interacting) instantons with the same $\tau$,\cite{Alvarez2013aboveTc} will be the subject of future work.

\subsubsection*{Understanding the theory}
Before demonstrating the accuracy of the theory numerically, we begin %
with a  qualitative discussion of the terms that appear and how they interact with one another.
Perhaps the first thing one notices, is that the $n$-orbit terms appear with alternating signs. In the  derivation from WKB presented in the Appendix, these alternating signs occur as a direct consequence of the form of the uniform transmission probability. However, we note that the necessity of the sign alternation is also evident from Eqs.~(\ref{eq:First_Key_Result_2a}) and (\ref{eq:First_Key_Result_2})
as it is required to match the alternating divergences of $(\sin(\omega\tau/2))^{-1}$ in $k_{\text{m-pb}}$. To understand this cancellation explicitly, we consider the behaviour of $k_{\text{m-pb}}$ about $\tau\to n\tau_{\rm c}$. Using the following expansion around $\tau=n\tau_{\rm c}$
\begin{equation}
    \frac{\omega}{4\pi \sin(\tau\omega/2)} \sim \frac{(-1)^n}{2\pi(\tau-n\tau_{\rm c})} +\mathcal{O}(\tau-n\tau_{\rm c})
\end{equation}
one can show that the parabolic barrier rate behaves like
\begin{equation}
    e^{\tau V^{\ddagger}/\hbar} k_{\text{m-pb}}Z_r\sim \frac{Z^\ddagger(n\tau_{\rm c})(-1)^n}{2\pi (\tau-n\tau_{\rm c})} +\frac{{Z^{\ddagger}}'(n\tau_{\rm c})(-1)^n}{2\pi} , \label{eq:Divergence_1}
\end{equation}
again with an error of $\mathcal{O}(\tau-n\tau_{\rm c})$.
To see how this divergent behaviour is cancelled, we expand the terms appearing in the first sum of Eq.~(\ref{eq:Full_Uniform_Theory_simplified}) about $\tau=n\tau_{\rm c}$, retaining terms up to $\mathcal{O}(\tau-n\tau_{\rm c})$, to give
\begin{equation}
    \frac{\mathrm{sgn}\!\left(\tau-n\tau_{\rm c}\right)}{\sqrt{4\pi\Delta S_{\!n}/\hbar}}  \sim \sqrt{\frac{n\hbar/2\pi}{- S''_{\!\rm inst}(\tau_{\rm c})}} \left(\frac{1}{\tau-n\tau_{\rm c}}-\frac{S'''_{\rm inst}(\tau_{\rm c})}{6n S''_{\rm inst}(\tau_{\rm c})}\right).
\end{equation}
Using this result, and noting that the instanton partition function is related to the usual transition state partition function according to $Z_{n,\mathrm{inst}}(n\tau_{\rm c})=Z^\ddagger(n\tau_{\mathrm{c}})$, it can then be shown that
\begin{equation}
\begin{aligned}
    \mathrm{sgn}\!\left(\tau-n\tau_{\rm c}\right) \frac{k_{n,\rm inst}Z_re^{+S_{\!n,\mathrm{inst}}/\hbar}}{\sqrt{4\pi\Delta S_{\!n}/\hbar}} &\sim \frac{Z^\ddagger(n\tau_{\rm c})}{2\pi (\tau-n\tau_{\rm c})}\\ &+ \frac{Z^\ddagger(n\tau_{\mathrm{c}})}{2\pi}\frac{S'''_{\rm inst}(\tau_{\rm c})}{3n S''_{\rm inst}(\tau_{\rm c})}\\ &+ \frac{Z_{n,\mathrm{inst}}'(n\tau_{\mathrm{c}})}{2\pi} \label{eq:Divergence_2}
\end{aligned}
\end{equation}
again to $\mathcal{O}(\tau-n\tau_{\rm c})$.
Comparing Eq.~(\ref{eq:Divergence_1}) and Eq.~(\ref{eq:Divergence_2}) we see that [once we combine  Eq.~(\ref{eq:Divergence_2}) with the factor of $(-1)^{n+1}$] the divergent terms exactly cancel leaving just a constant as $\tau\to n\tau_c$. Hence, exactly at the crossover temperature, our theory predicts a correction to the rough ``factor of 2 error'' of instanton theory discussed earlier. Specifically, (ignoring the hyperasymptotic multi-orbit terms) we find that 
\begin{equation}
\begin{aligned}
    k(\tau_c)- \frac{1}{2}k_{\rm inst}(\tau_{\rm c})  \sim  \frac{1}{2\pi}e^{-\tau_{\rm c} V^\ddagger/\hbar} \Bigg(& \frac{Z^\ddagger(\tau_{\mathrm{c}})}{Z_r(\tau_{\mathrm{c}})}\frac{S'''_{\rm inst}(\tau_{\rm c})}{3S''_{\rm inst}(\tau_{\rm c})}\\
+&\frac{Z_{\rm inst}'(\tau_{\mathrm{c}})}{Z_r(\tau_{\mathrm{c}})}-\frac{{Z^\ddagger}'(\tau_{\mathrm{c}})}{Z_r(\tau_{\mathrm{c}})} \Bigg). \label{eq:error_correction}
    \end{aligned}
\end{equation}

Having considered the behaviour of the theory close the crossover temperature(s) (where $\Delta S_{\!n}=0$), let us now consider the behaviour of the new terms for $\Delta S_{\!n}/\hbar \gg 1$, both when $\tau>n\tau_{\rm c} $ and $\tau<n\tau_{\rm c} $. The simpler of these two cases is $\tau>n\tau_{\rm c} $, where (half) the complementary error function approaches one and hence we have
\begin{equation}
\begin{aligned}
     \frac{k_{n,\rm inst}e^{-\Delta S_{\!n}/\hbar}}{\sqrt{4\pi\Delta S_{\!n}/\hbar}}  + k_{n,\rm inst}\, \frac{1}{2}\mathrm{erfc}\!\left(-\sqrt{\Delta S_{\!n}/\hbar}\right)  \sim  k_{n,\rm inst}. 
\end{aligned}
\end{equation}
The slightly more complicated case is $\tau<n\tau_{\rm c} $. Here we can make use of the standard asymptotic result
\begin{equation}
    \mathrm{erfc}\!\left(x\right)\sim \frac{e^{-x^2}}{\sqrt{\pi x^2}}\left(1-\frac{1}{2x^2}+\dots\right)\text{ as $x\to\infty$}
\end{equation}
from which we observe that
\begin{equation}
    \!\!-\frac{k_{n,\rm inst}e^{-\Delta S_{\!n}/\hbar}}{\sqrt{4\pi\Delta S_{\!n}/\hbar}}  + k_{n,\rm inst}\, \frac{1}{2}\mathrm{erfc}\!\left(\!\!\sqrt{\Delta S_{\!n}/\hbar}\right)  \sim -\frac{k_{n,\rm inst}e^{-\Delta S_{\!n}/\hbar}}{4\!\sqrt{\pi(\Delta S_{\!n}/\hbar)^3}} \!
\end{equation}
which as we would expect is subdominant to $k_{\text{m-pb}}$ even for $n=1$.

The derivation from the WKB transmission probability given in the Appendix suggests a natural separation of the theory into three parts: classical above barrier transmission, quantum above barrier reflection, and quantum tunnelling. Making this separation we can write Eq.~(\ref{eq:Full_Uniform_Theory_simplified}) as
\begin{equation}
    k \sim k_{\rm TST} - k_{\rm reflect} + k_{\rm tunnel}
\end{equation}
where
\begin{equation}
    k_{\rm TST} = \frac{1}{2\pi\tau}\frac{Z^\ddagger(\tau)}{Z_r(\tau)}e^{-\tau V^\ddagger/\hbar}
\end{equation}
is the above barrier transmission contribution (equivalent to the TST rate with $\kappa=1$). The reflection rate can then be expressed as\footnote{Note that the sum in this expression can be expressed in terms of the Lerch transendent special function.}  
\begin{equation}
    k_{\rm reflect} = \frac{1}{2\pi}\frac{Z^\ddagger(\tau)}{Z_r(\tau)}e^{-\tau V^\ddagger/\hbar}\sum_{\lambda=1}^\infty (-1)^{\lambda+1}\frac{1}{\lambda \tau_{\rm c}+\tau} \label{eq:k_reflect}
\end{equation}
where to avoid potential confusion in the following sections we have used $\lambda$ rather than $n$ as the dummy index in the sum.
Note that as defined $k_{\rm reflect}$ is always positive, and can be expressed as $k_{\rm reflect}=\phi(\tau/\tau_{\rm c})k_{\rm TST}$ where $\phi(x)$ is a system independent function, with $\phi(0)=0$, $\phi(1)=1-\ln(2)$, and $\phi(\infty)=\nicefrac{1}{2}$. Finally the tunneling contribution, which is system dependent can be expressed as
\begin{equation}
\begin{aligned}
    k_{\rm tunnel}&= \frac{1}{2\pi}\frac{Z^\ddagger(\tau)}{Z_r(\tau)}e^{-\tau V^\ddagger/\hbar}\sum_{\lambda=1}^\infty (-1)^{\lambda+1}\frac{1}{\lambda \tau_{\rm c}-\tau} \\&+ \sum_{n=1}^\infty (-1)^{n+1} \mathrm{sgn}\!\left(\tau-n\tau_{\rm c}\right) \frac{k_{n,\rm inst}e^{-\Delta S_{\!n}/\hbar}}{\sqrt{4\pi\Delta S_{\!n}/\hbar}}  \\
    &+\sum_{n=1}^\infty (-1)^{n+1} k_{n,\rm inst}\, \frac{1}{2}\mathrm{erfc}\!\left(\mathrm{sgn}\!\left(n\tau_{\rm c}-\tau\right)\!\sqrt{\frac{\Delta S_{\! n}}{\hbar}}\right).
    \end{aligned} \label{eq:k_tunnel}
\end{equation}

\subsubsection*{Numerical Considerations}
Having given a qualitative interpretation of the theory we now turn to some practical considerations about the numerical implementation of the theory. First, the presence of infinite sums in Eq.~(\ref{eq:Full_Uniform_Theory_simplified}) might appear daunting, and raises the obvious question: How many terms are needed to reach numerical conversion? Clearly, one requires  enough terms to avoid the divergence of $k_{\text{m-pb}}$ at the temperature of interest. We will see in the next section that this is the dominant consideration and that for $0<\tau<2\tau_{\rm c}$ excellent convergence is obtained with only $n=2$. It should be stressed that if implementing the theory using Eqs.~(\ref{eq:k_reflect}) and (\ref{eq:k_tunnel}) one must include a sufficient number of terms in the sum over $\lambda$ to recover the parabolic barrier rate above the crossover temperature, and hence the maximum value of $\lambda$ may be higher than $n$. Note that including an arbitrarily large number of terms in this sum is trivial.

One remaining aspect that we have not yet addressed is the meaning of $S_{\!\rm inst}(\tau)$ [and $Z_{\rm inst}(\tau)$] when $\tau<\tau_{\rm c}$.
Because there exist no real-position periodic orbits for these values of imaginary time, the resulting trajectories must move into the complex position plane. Finding such trajectories is clearly impractical for realistic chemical applications. However, since the terms containing $S_{\!\rm inst}(\tau<\tau_{\rm c})$ are always subdominant we can develop numerical approximations of the action in this region that require only real positions without affecting the key asymptotic behaviour of the theory.
In the following section we will give an example of how this can be for one-dimensional systems, and compare to the results obtained using the exact action.

\section{Numerical Results}\label{sec:results}
\subsection{Symmetric Eckart Barrier}
To illustrate the accuracy of the new theory we consider the prototypical one-dimensional model of reactive scattering: the symmetric Eckart barrier. The potential for the symmetric Eckart barrier is defined as
\begin{equation}
  V(q) = V^\ddagger \sech^2(q/L). 
\end{equation}
For this simple model system the instanton action can be evaluated analytically as
\begin{equation}
    S_{\!\rm inst}(\tau) = 2  \tau_{\rm c}V^\ddagger - \frac{ \tau_{\rm c}^2 V^\ddagger}{\tau }
\end{equation}
where the barrier frequency is given by
\begin{equation}
    \omega = \sqrt{\frac{2 V^\ddagger}{m L^2}}.
\end{equation}
To aid comparison with previous work\cite{Voth+Chandler+Miller,Voth1989correlation,RPMDrefinedRate,Richardson2009RPInst,Faraday,Upadhyayula2023UniformInstanton} we consider the following parameters: $L=0.66047\, a_0$, $m=1836\,m_e $, $V^\ddagger=72\hbar^2/(m \pi^2  L^2)$. The exact result was calculated for comparison by numerical integration of 
\begin{equation}
    k Z_r = \frac{1}{2\pi\hbar}\int_0^\infty  P(E)e^{-\tau E/\hbar} \mathrm{d}E \label{eq:exact_PE_integrated}
\end{equation}
using the exact analytical result for the transmission probability, $P(E)$.\cite{Eckart,BellBook}

\begin{figure}[t]
    \centering
    \includegraphics[width=1.0\linewidth]{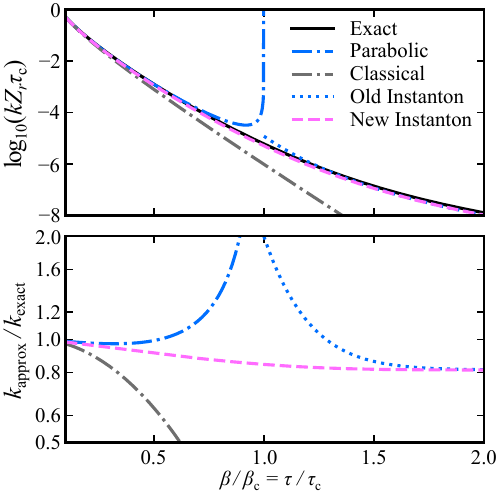}
    \caption{Rates and ratio of approximate and exact rate for the symmetric Eckart barrier as a function of inverse temperature. The ratio of approximate to exact results in the lower panel is plotted on a logarithmic scale.}
    \label{fig:symmetric_eckart_results}
\end{figure}

Figure \ref{fig:symmetric_eckart_results} compares the rate calculated with the new theory against the exact, classical, parabolic barrier, and standard (old) instanton results for the symmetric Eckart barrier.  
The top panel shows the logarithm of the rate for each theory (multiplied by $Z_r\tau_{\rm c}$ to ensure it is dimensionless), and the bottom panel shows the ratio of the approximate theories to the exact result (plotted on a logarithmic scale so that overestimation and underestimation are treated on an equal footing).
One immediately notices the divergence of the parabolic barrier rate at the crossover temperature ($\beta/\beta_{\rm c}=1$) and the (approximate) factor of two error of the standard instanton result.

In contrast to both the standard theory and parabolic barrier rate, the new theory smoothly connects the low and high temperature limits. The accuracy of the new theory is as good as one could have hoped, with the error no greater in the vicinity of the crossover temperature than it is at low temperature (where it is approximately 20\% too small). It is important to stress that quantum effects are not insignificant in this regime. The quantum rate is more than a factor of 6 larger than the classical rate at $T_{\rm c}$ and still more than a factor of 2 larger than the classical rate even at  $T=1.5\times T_{\rm c}$. This highlights the importance of the new theory in accurately describing the onset of quantum tunneling.

At lower temperatures, we see that the error of both the new theory and the standard instanton have approximately plateaued at $\tau=2\tau_{\rm c}$ (half the crossover temperature). The close agreement between the standard theory and the new theory at $\tau=2\tau_{\rm c}$ indicates that in this regime the multi-orbit terms do not significantly affect the rate. However, it is important to remember that if one were to neglect these terms and try to use Eq.~(\ref{eq:First_Key_Result_2}) instead of Eq.~(\ref{eq:Full_Uniform_Theory_simplified}) the result would diverge at $\tau=2\tau_{\rm c}$. One could of course consider approximating the full theory [Eq.~(\ref{eq:Full_Uniform_Theory_simplified})] by removing both the multi-orbit terms and making appropriate modifications to $k_{\text{m-pb}}$ to remove the divergences. This would break the formal asymptotic properties of the theory, but may nevertheless provide a useful approximation to the rate. 

Of course, while the present method solves the breakdown of instanton theory at the crossover temperature it is still an approximate theory. To obtain more accurate results one would need to include higher order terms in the asymptotic series. This idea has been explored in a recent paper in the context of the calculation of ground state tunneling splittings.\cite{Lawrence2023RPI+PC}  There, it was shown that the ring-polymer instanton (RPI) approach could be improved by calculating higher order terms in the asymptotic series, which can be thought of as anharmonic perturbative corrections. The resulting perturbatively corrected theory (RPI+PC) was shown to be a practical method for real molecules with application to malonaldehyde,\cite{Lawrence2023RPI+PC} requiring the calculation of third and fourth derivatives of the potential along the instanton. The resulting theory therefore shares similarities with vibrational perturbation theory (VPT2) and semiclassical transition state theory (SCTST), which involve third and fourth derivatives of the potential at the transition state.\cite{Miller1990SCTST,Nguyen2010SCTST,Wagner2013SCTST,Conte2024PerspectiveSCTST_and_Spectra} Future work will look to extend RPI+PC to the calculation of reaction rates, which in combination with the present theory should provide an alternative to the SCTST formalism. It should be noted that inspired by SCTST, Upadhyayula and Pollak have developed a theory that incorporates anharmonic corrections within their approach to instanton theory by adding an anharmonic $\hbar$ dependent (but temperature and energy independent) correction to the action.\cite{Upadhyayula2024hbar2corrections,Pollak2024hbar4corrections,Pollak2024Perspective} A detailed discussion of the connections and differences between their perspective and the present framework is left for future work.

\subsection{Asymmetric Eckart Barrier}

Next we consider the asymmetric version of the Eckart barrier, for which the potential can be written as
\begin{equation}
    V(q) = \frac{\left(\sqrt{V_1}+\sqrt{V_2}\right)^2}{4\cosh^2(q/L)} -\frac{V_2-V_1}{1+\exp(-2q/L)},
\end{equation}
note that $V_1=V^\ddagger$.
For this system evaluating the action analytically is slightly more involved. Rather than give the expression for $S_{\! \rm inst}(\tau)$ explicitly we  instead note it can be computed from knowledge of the reduced action, $W(E)$. Note the reduced action is the Legendre transform of the action, $S_{\! \rm inst}(\tau)=W(E)+\tau E$. %
The reduced action can be expressed as\cite{Upadhyayula2023UniformInstanton}
\begin{equation}
    W(E) = A  \left( 1+\sqrt{\frac{V_2}{V_1}} - \sqrt{\frac{E}{V_1}} - \sqrt{\frac{E}{V_1}-1+\frac{V_2}{V_1}}\right)
\end{equation}
where
\begin{equation}
    A = \frac{4\pi}{1+\sqrt{V_2/V_1}}\sqrt{\frac{V_1 V_2}{\omega^2}}.
\end{equation}
One can then obtain $S_{\! \rm inst}(\tau)$, by noting that the period of the orbit, $\tau$, is related to the energy by $\tau(E)=W'(E)$. Inverting this to find $E(\tau)$ then allows one to calculate $S_{\! \rm inst}(\tau)$.

While for these simple models one can obtain analytical expressions for the action, in general it must be found numerically e.g.~using the ring-polymer instanton method.\cite{Andersson2009Hmethane,Richardson2009RPInst,Perspective,Richardson2018InstReview} For $\tau>\tau_{\rm c}$ this is precisely what is already done in standard instanton calculations. 
However, for $\tau<\tau_{\rm c}$ there does not exist a real periodic orbit. One might reasonably be concerned that finding a trajectory in complex positions would be impractical, however, we will now argue that this is unnecessary. 
Note first that the terms in Eq.~(\ref{eq:Full_Uniform_Theory_simplified}) containing $S_{\!\rm inst}(\tau)$ for $\tau<\tau_{\rm c}$ are subdominant.
This allows us to approximate the action in this region without changing the key asymptotic properties of the theory.  
For example, when $S'''_{\!\rm inst}(\tau_{\rm c})>0$, we can use 
\begin{equation}
    S_{\!\rm inst}(\tau) \approx \begin{cases} \begin{array}{ll} \displaystyle S_{\!\rm inst}(\tau)\vphantom{\sum_{j=1}^3} & \text{if } \tau\geq\tau_{\rm c} \\ \displaystyle S_{\!\rm inst}(\tau_{\rm c}) + \frac{1}{\tau}\sum_{j=1}^3   \frac{s_j}{j!}(\tau-\tau_{\rm c})^j & \text{if } \tau<\tau_{\rm c} \label{eq:action_approx}
\end{array}
\end{cases}
\end{equation}
where
\begin{equation}
\begin{aligned}
    s_1 &= S'_{\!\rm inst}(\tau_{\rm c})\tau_{\rm c}\equiv V^\ddagger \tau_{\rm c} \equiv S_{\!\rm inst}(\tau_{\rm c}) \\
    s_2 &= 2S'_{\!\rm inst}(\tau_{\rm c})+S''_{\!\rm inst}(\tau_{\rm c})\tau_{\rm c} \\
    s_3 &= 3S''_{\!\rm inst}(\tau_{\rm c}) + S'''_{\!\rm inst}(\tau_{\rm c})\tau_{\rm c}.
    \end{aligned}
\end{equation}
Under the condition that $S'''_{\!\rm inst}(\tau_{\rm c})>0$, this approximation satisfies a number of key properties. First, it matches the first three derivatives of the action at $\tau_{\rm c}$. Second, it correctly predicts that $S_{\! \rm inst}(\tau)\to-\infty$ as $\tau\to0$. Finally, it also satisfies $S_{\! \rm inst}(\tau)<\tau V^\ddagger$, and hence it is guaranteed that $\Delta S_{\! n}>0$. Importantly, as this approximation only involves information at $\tau\geq\tau_{\rm c}$ it is straightforward to evaluate using standard techniques.  For example, $S''(\tau)$ can be calculated directly from a ring-polymer instanton using the identities given in Refs.~\citenum{Richardson2018InstReview} and \citenum{GoldenRPI}, from which $S'''(\tau_{\rm c})$ can be calculated by finite difference. We have considered here only the case of $S'''(\tau_{\rm c})>0$, as this is expected to be the usual situation for an instanton that collapses smoothly to the transition state, alternative approaches for systems where $S'''(\tau_{\rm c})<0$ will be considered in future work.

As with the symmetric Eckart barrier the parameters for the model are chosen to correspond to previous studies\cite{Voth+Chandler+Miller,Voth1989correlation,RPMDrefinedRate,Richardson2009RPInst,Upadhyayula2023UniformInstanton} and are given in reduced units as
$m=1$, $L=8/\sqrt{3\pi}$, 
$V_1=(6/\pi)^{\nicefrac{1}{4}}$, 
$V_2=(24/\pi)^{\nicefrac{1}{4}}$, 
and $\hbar=1$. The exact results are again calculated using Eq.~(\ref{eq:exact_PE_integrated}), with the exact analytical $P(E)$.\cite{Eckart,Miller1979unimolecular}  

\begin{figure}[t]
    \centering
    \includegraphics[width=1.0\linewidth]{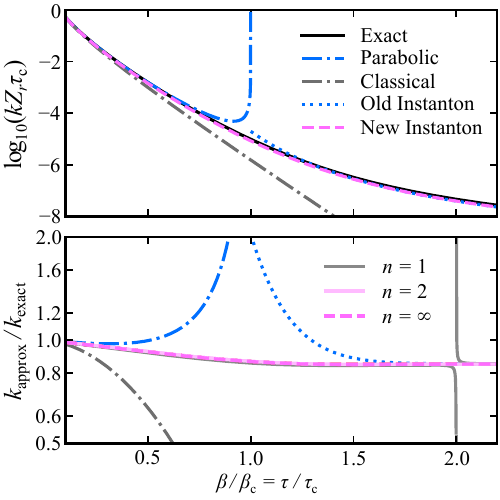}
    \caption{Rates and ratio of approximate and exact rate for the asymmetric Eckart barrier as a function of inverse temperature. The ratio of approximate to exact results in the lower panel is plotted on a logarithmic scale. The lower panel also illustrates the convergence with respect to the maximum values of $n$ in the sum in Eq.~(\ref{eq:Full_Uniform_Theory_simplified}) The results for the new instanton method using the exact $S_{\!\rm inst}(\tau)$ or Eq.~(\ref{eq:action_approx}) are graphically indistinguishable. }
    \label{fig:asymmetric_eckart_results}
\end{figure}

Figure \ref{fig:asymmetric_eckart_results} compares the rate calculated with the new theory against the exact, classical, parabolic barrier, and standard (old) instanton results for the asymmetric Eckart barrier. We see that the new theory performs just as well for the asymmetric barrier as it does for the symmetric one. This is an important property of the standard instanton result that the present theory also retains, and stands in contrast to centroid based path-integral methods that are known to breakdown at low temperatures for asymmetric systems.\cite{Voth+Chandler+Miller,Voth1989correlation,RPMDrefinedRate,Richardson2009RPInst} Pleasingly, we find that the results of the new theory using either the exact action, or the approximate action given by Eq.~(\ref{eq:action_approx}) are graphically indistinguishable at all values of $\tau$ for this system, differing by at most $0.02\%$. This illustrates that the new theory can be accurately applied using just information that is available from standard ring-polymer instanton calculations. 

 The lower panel of Fig.~\ref{fig:asymmetric_eckart_results} also shows how the results converge with increasing number of instanton orbits included in the sums in Eq.~(\ref{eq:Full_Uniform_Theory_simplified})\@. One can immediately see that, for this range of temperatures, including terms up to $n=2$ already agrees almost perfectly with the full sum: the largest deviation being just over 0.2\%. In fact, even including only a single orbit ($n=1$) is practically sufficient for nearly all temperatures considered, with the only significant deviation occurring in a very narrow region near the divergence at $\beta/\beta_{\rm c}=2$. Away from this divergence the largest deviation occurs close to the crossover temperature, where the full sum is just over 1\% larger than the single orbit result. This is a significant result as it indicates that Eq.~(\ref{eq:First_Key_Result_2}), which only involves information available from a standard instanton calculation, is already sufficient for studying the behaviour of the rate for a wide range of temperatures near crossover. Finally, the narrow range of temperatures affected by the divergence at $\beta/\beta_{\rm c}=2$ may seem surprising when compared to the broad divergence of the parabolic barrier rate. However, this difference can be understood by noting that, in the case of the parabolic barrier rate, the diverging term is asymptotically dominant, whereas at $\beta/\beta_{\rm c}=2$ the term that is divergent is formally subdominant and has to overcome an exponential suppression.

\section{Discussion}\label{sec:discussion}
Despite the limited numerical significance of the multi-orbit terms for the systems studied here, their natural appearance in the theory is theoretically interesting. For a fixed value of $\tau$ these multi-orbit terms are formally subdominant (i.e.~negligible in comparison) to every term in the asymptotic series for the single orbit. Therefore, they do not contribute within a basic asymptotic analysis, be it either Poincar\'e asymptotics, where a fixed number of terms in the asymptotic series are included, or superasymptotics, where the number of terms is chosen to minimise the error.\cite{Berry1990Hyperasymptotics} The multi-orbit terms, instead, correspond to what are often referred to as hyperasymptotic contributions.\cite{Berry1990Hyperasymptotics}
These terms are significant because
the error made by the original asymptotic series can be written recursively in terms of them.
Hyperasymptotic series and their resummation are a central part of the closely related areas of resurgent asymptotics, exponential asymptotics, and transseries.\cite{Dingle1973AsymptoticsBook,Ecalle1981Resurgence,Dunne2014ResurgentTransseries,Aniceto2019PrimerResurgentTransseries} Such approaches allow one to go beyond the accuracy of superasymptotics and in principle to even obtain exact results. While these techniques have found a wide range of uses in modern physics, with applications in string theory, quantum field theory, and cosmology,\cite{Aniceto2019PrimerResurgentTransseries} they have as yet not found wide use in chemistry or chemical physics. 
It will therefore be interesting in the future to make connection to these approaches, for example by using the method proposed by Berry and Howls\cite{Berry1991Hyperasymptotics} to derive the multi-orbit contributions to the theory.

It should be noted that Miller's original derivation of the standard thermal instanton theory proceeded via a microcanonical expression involving multi-orbit instantons.\cite{Miller1975semiclassical} This raises the obvious question, how is the present theory related to Miller's original result?
While Miller's microcanonical result is equivalent in one dimension to the uniform WKB transmission probability, Miller noted immediately that it did not recover the correct result in a separable system. Despite this, once integrated asymptotically over energy it recovers the thermal instanton, which does correctly describe both separable and non-separable systems.
One might, therefore, wonder, can Eq.~(\ref{eq:Full_Uniform_Theory_simplified}) be derived from Miller's microcanonical expression following the method used in the Appendix in one dimension. 
Interestingly, the answer is no. One finds that, the resulting theory is incorrect, failing to recover the separable result except at low temperature.

Miller's original expression is not the only microcanonical instanton theory. Shortly after his first paper on instanton theory, Miller, along with Chapman and Garrett,  suggested an ad-hoc correction to the original microcanonical formula designed to correctly recover the separable limit.\cite{Chapman1975rates} 
More recently an alternative (easier to implement) method, known as the density of states microcanonical instanton, has been proposed.\cite{DoSTMI,JoeFaraday}
Notably, thermalising either of these microcanonical theories  by integrating asymptotically ($\hbar\to0$) over energy does recover Eq.~(\ref{eq:Full_Uniform_Theory_simplified}).\footnote{This is trivially true starting from the density of states instanton. The connection to Chapman, Garrett, and Miller's expression can be made by noting that it differs from the density of states instanton by a term of order $\hbar$ that can, therefore, be discarded.} 
It is important to stress, however, that the present theory should not be considered as an approximation to these microcanonical results. This can be seen most clearly for escape from a metastable well, for which the thermal instanton correctly describes the plateau in the rate at low temperature, whereas exact integration of the semiclassical transmission probability approaches zero.  
It should also be emphasised that, for the calculation of thermal rates the present approach is also more practical than thermalising a microcanonical theory.\cite{Faraday,DoSTMI}
This is because integrating over energy requires information from instanton calculations at a wide range of imaginary times, including instantons whose period is greater than the thermal time. In contrast, Eq.~(\ref{eq:Full_Uniform_Theory_simplified}) only requires calculations at the temperature of interest, $T$, and a small number of integer multiples, $nT$.

Instantons are, of course, not the only way to incorporate the effects of tunneling into the calculation of reaction rate constants. In particular path-integral theories, such as ring-polymer molecular dynamics (RPMD)\cite{RPMDrate,RPMDrefinedRate,Suleimanov2013RPMDrate,Lawrence2020rates} and related quantum transition state theories,\cite{Richardson2009RPInst,Cao1996QTST,Hele2013QTST} are capable of describing the transition between the high and low temperature regimes.
However, it is important to recognise that, as these approaches involve path-integral sampling, they are practically very different methods.
In particular, numerical determination of free-energy differences, and the associated need to compute the potential energy at a large number of configurations typically makes sampling methods more expensive than instanton theory. The present theory and RPMD rate theory thus have different use cases. RPMD is most useful in liquid systems where instanton theory cannot be applied,\cite{RPMDprotonTransfer,Boekelheide2011DHFR} and instanton theory is most useful for gas phase and surface reactions in combination with high-level ab-initio electronic structure theory.\cite{GPR,Fang2024CheapInstantons,HCH4,Meisner2011isotope,Cooper2018interpolation,McConnell2019instanton,Litman2022InstantonElectronicFrictionII} 

One might reasonably suggest that making a harmonic approximation to RPMD-TST would avoid the need to perform path-integral sampling, and hence give a competitor to the present theory. However, to evaluate harmonic RPMD-TST in the vicinity of the crossover temperature one would need to derive a uniform theory that would likely be very similar to Eq.~(\ref{eq:Full_Uniform_Theory_simplified}). This is because, above the crossover temperature a basic harmonic approximation to RPMD-TST recovers the parabolic barrier rate,\cite{RPMDrate} and below the crossover temperature it is closely related to the $\Im F$ formulation of the thermal instanton.\cite{Richardson2009RPInst} %

\section{Conclusions and future work}\label{sec:conclusion}
The breakdown of instanton theory at the crossover temperature has been a long standing problem in reaction rate theory. Although suggestions had been made to overcome this problem,\cite{Affleck1981ImF,Grabert1984Crossover,Hanggi1986ImF,Cao1996QTST,Kryvohuz2011rate,McConnell2017instanton,Upadhyayula2023UniformInstanton} none were entirely satisfactory. Here the problem has been  solved using rigorous asymptotic analysis for the general class of multidimensional problems in which the instanton collapses smoothly to the transition state.
To derive this result we have combined a new real-time version of Richardson's flux-correlation function derivation of instanton theory\cite{Richardson2016FirstPrinciples,Richardson2018InstReview} with the modern method for developing uniform asymptotic expansions due to Bleistein.\cite{Bleistein1966Uniform}
The resulting theory is a rigorous semiclassical theory for the rate that is uniformly valid as a function of $\tau=\beta\hbar$. Unlike previous approaches\cite{Affleck1981ImF,Grabert1984Crossover,Hanggi1986ImF,Cao1996QTST,Kryvohuz2011rate,McConnell2017instanton} the present result is a smooth function of temperature.
The new theory is also rigorously independent of dividing surface, and obeys detailed balance. 
Although the derivation was motivated by considering real-time trajectories, the final result only involves imaginary-time quantities (just like the original instanton theory).
Importantly, close to the crossover temperature the new theory only requires information already available from a standard instanton calculation, meaning that the theory can immediately be applied within existing instanton codes. 

There are a number of exciting avenues for further the theoretical development. First, the present theory is an important step towards the development of more accurate microcanonical instanton theories. In particular, as the present theory is a globally valid (and smooth) function of temperature, it is now possible to use the steepest descent approximation to the inverse Laplace transform to obtain the microcanonical rate at any energy.\cite{Forst1971InverseLaplaceTransform,Tao2020microcanonical,DoSTMI,JoeFaraday} [Note in this approach the asymptotic parameter is \emph{not} associated with $\hbar$.] 
The real-time formalism developed here also opens up exciting opportunities for future research. Although the real-time components of the trajectories do not contribute to the current theory, they could be used to extract extra information in other contexts, such as vibrationally state-resolved or electronically nonadiabatic reaction rates.
Another interesting direction for further development lies in connecting the present results to the $\Im F$ formulation. While the derivation used here was based on reactive scattering, the final result should be equally applicable to escape from a metastable well---the basis of the $\Im F$ derivation of instanton theory. Exploring this connection further would nevertheless be useful to link the current work more closely to the high-energy physics literature.

A key advantage of the first-principles derivation of the new theory is that it provides a rigorous framework for future work.
One such area is the generalisation of the theory to more complex systems in which the instanton does not collapse smoothly to the transition state with increasing temperature. 
While these systems have not received much attention so far in the chemical literature, this is likely because there has not existed a theory that can treat them.
Another advantage of the present theory is that it can be systematically improved via the inclusion of higher order terms in the asymptotic series. These perturbative corrections have already been incorporated into the ring-polymer instanton (RPI) formalism (to give RPI+PC) for the calculation of molecular tunneling splittings.\cite{Lawrence2023RPI+PC}
Future work will look to implement these corrections in the context of reaction rates to help describe systems where there is a significant change in anharmonicity along the reaction coordinate.
Such a combination of RPI+PC with the present theory would be a strong competitor to SCTST,\cite{Miller1990SCTST,Nguyen2010SCTST,Wagner2013SCTST,Conte2024PerspectiveSCTST_and_Spectra} with the important advantage of providing a rigorous description of deep tunneling.    

\section*{Supplementary Material}
The supplementary material contains derivations of Eqs.~(\ref{eq:action_at_infinity}) and (\ref{eq:prefactor_at_infinity}). It also contains a comparison of the present approach to the recent proposal of Upadhyayula and Pollak that illustrates that their result is not a rigorous uniform asymptotic approximation to the rate constant in $\tau$ as $\hbar\to0$.

\section*{Data Availability Statement}
The data that support the findings of this study are available from the corresponding author upon reasonable request. 

\section*{Acknowledgements}
I would like to thank Jeremy Richardson and George Trenins for helpful discussions. 
This work was supported by an Independent Postdoctoral Fellowship at the Simons Center for Computational Physical Chemistry, under a grant from the Simons Foundation (839534, MT).

\section*{Appendix: Derivation of main result in one dimension from Kemble's uniform semiclassical transmission probability}\label{App:Energy_Derivation}
\renewcommand{\theequation}{A\arabic{equation}}
\setcounter{equation}{0}
In the main text we derive our uniform expression in the time domain, using methods for the asymptotic evaluation of integrals.
In one dimension an alternative but equivalent approach is to use WKB analysis. These two approaches are equivalent as the asymptotic parameter is formally the same. To arrive at a uniform expression for the thermal rate we can, therefore, begin with the uniform WKB expression for the transmission probability as a function of energy
\begin{equation}
    P_{\rm SC}(E) = \frac{1}{1+e^{W(E)/\hbar}}. \label{eq:Kemble_P_SC}
\end{equation}
This expression was originally proposed by Kemble\cite{Kemble1935WKB} in 1935 and was later rigorously derived by Fr\"oman and Fr\"oman.\cite{Froman1965JWKB} 
It involves the reduced Euclidean action, $W(E)$, which is related to $S_{\!\rm inst}(\tau)$ via a Legendre transform as
\begin{equation}
    S_{\!\rm inst}(\tau) = W(E) + \tau E
\end{equation}
with
\begin{equation}
    \tau = -W'(E)
\end{equation}
and
\begin{equation}
    E = S_{\!\rm inst}'(\tau).
\end{equation}

When $E>V^\ddagger$, then $W(E)<0$ and we can write
\begin{equation}
    P_{\rm SC}(E) = \sum_{n=0}^\infty (-1)^n \,e^{\,n W(E)/\hbar}. \label{eq:Above_Barrier_Expansion}
\end{equation}
Similarly when $E<V^\ddagger$ then $W(E)>0$ and we can write 
\begin{equation}
    P_{\rm SC}(E) = e^{- W(E)/\hbar}\sum_{n=0}^\infty (-1)^n \, e^{-n W(E)/\hbar}. \label{eq:below_barrier_expansion}
\end{equation}

To obtain the correct uniform asymptotic expression for the thermal rate we therefore begin by writing
\begin{equation}
    k Z_r \sim \frac{1}{2\pi\hbar} \int_0^\infty P_{\rm SC}(E) e^{-\tau E/\hbar} \mathrm{d}E.  
\end{equation}
Then separating this into two parts
\begin{equation}
\begin{aligned}
    kZ_r\sim\frac{1}{2\pi\hbar}& \int_0^{V^\ddagger} P_{\rm SC}(E) e^{-\tau E/\hbar} \mathrm{d}E \\
    +\frac{1}{2\pi\hbar} &\int_{V^\ddagger}^\infty P_{\rm SC}(E) e^{-\tau E/\hbar} \mathrm{d}E
\end{aligned}
\end{equation}
allows us to insert the expansions Eqs.~(\ref{eq:Above_Barrier_Expansion}) and (\ref{eq:below_barrier_expansion}) to give
\begin{equation}
\begin{aligned}
    kZ_r\sim\frac{1}{2\pi\hbar}& \int_0^{V^\ddagger} \sum_{n=1}^\infty (-1)^{n+1} \, e^{-n W(E)/\hbar-\tau E/\hbar} \mathrm{d}E \\
    +\frac{1}{2\pi\hbar} &\int_{V^\ddagger}^\infty \sum_{n=0}^\infty (-1)^n \,e^{\,n W(E)/\hbar-\tau E/\hbar} \mathrm{d}E.
\end{aligned}
\end{equation}

Consider first the integral up to the barrier height
\begin{equation}
  I_n(\hbar)=\frac{1}{2\pi\hbar}\int_0^{V^\ddagger} e^{-(n W(E)+\tau E)/\hbar} \mathrm{d}E.
\end{equation}
Integrating by steepest descent gives the stationary condition as
\begin{equation}
    n W'(E_n^\star(\tau)) = -\tau.
\end{equation}
Now for $\tau/n<\tau_{\rm c}$ this stationary point will move outside of the integration range, $E_n^\star(\tau<n\tau_{\rm c})>V^\ddagger$. Hence to obtain a uniform expression valid when $E_n^\star>V^\ddagger$, we again make use of Bleistein's method.\cite{Bleistein1966Uniform,RWong1989UniformAsymptotics} Hence, defining
\begin{equation}
    S_{\!n,\mathrm{inst}}(\tau) = n W(E_n^\star)+\tau E_n^\star = n \,S_{\!\mathrm{inst}}(\tau/n)
\end{equation}
and
\begin{equation}
    \Delta S_{\!n} = \tau V^\ddagger - S_{\!n,\mathrm{inst}}(\tau)   
\end{equation}
application of Bleistein's method results in the following expression as $\hbar\to0$
\begin{equation}
\begin{aligned}
    I_n(\hbar)&\sim k_{n,\mathrm{inst}}Z_r\frac{1}{2}\mathrm{erfc}\!\left(\mathrm{sgn}\!\left(n\tau_{\rm c}-\tau\right)\sqrt{\frac{{\Delta S_{\!n}}}{{\hbar}}}\right) \\
    &+\frac{1}{2\pi}e^{-\tau V^\ddagger/\hbar} \frac{1}{n\tau_{\rm c} - \tau} \\
    &- \frac{1}{2\pi}e^{-\tau V^\ddagger/\hbar}\mathrm{sgn}\!\left(n\tau_{\rm c}-\tau\right) \sqrt{\frac{-S_{\!n,\mathrm{inst}}''(\tau)}{2\Delta S_{\!n}}} 
\end{aligned}
\end{equation}
where
\begin{equation}
    k_{n,\mathrm{inst}}Z_r = e^{-S_{\!n,\mathrm{inst}}(\tau)/\hbar} \frac{1}{\sqrt{2\pi\hbar}} \left(-\frac{\mathrm{d}^2 S_{\!n,\mathrm{inst}}}{\mathrm{d}\tau^2}\right)^{\!\nicefrac{1}{2}}. %
\end{equation}

Next we consider the integral above the barrier height
\begin{equation}
  J_n(\hbar)=\frac{1}{2\pi\hbar}\int_{V^\ddagger}^\infty e^{(n W(E)-\tau E)/\hbar} \mathrm{d}E.
\end{equation}
When evaluating this asymptotically, we note that $W'(E) < 0$. 
Hence, for $\tau>0$ then there is no stationary point inside the integration region and the integrand is peaked about $E=V^\ddagger$. The integral is, therefore,  approximated asymptotically by expanding the exponent linearly. Using,
$W'(V^\ddagger)=-\tau_{\rm c}=-2\pi/\omega$ we thus have
\begin{equation}
\begin{aligned}
    J_n(\hbar) &\sim \frac{1}{2\pi\hbar} \int_{V^\ddagger}^\infty e^{-\left(n\tau_{\rm c}(E-V^\ddagger)+\tau E\right)/\hbar} \mathrm{d}E \\
    & = \frac{1}{2\pi} e^{-\tau V^\ddagger/\hbar} \frac{1}{n\tau_{\rm c}+\tau}
\end{aligned}
\end{equation}
as $\hbar\to0$.

Combining these two results together we then obtain
\begin{equation}
    \begin{aligned}
        \!k Z_r&\sim \sum_{n=1}^\infty (-1)^{n+1} k_{n,\mathrm{inst}}Z_r\frac{1}{2}\mathrm{erfc}\!\left(\mathrm{sgn}\!\left(n\tau_{\rm c}-\tau\right)\!\sqrt{\frac{{\Delta S_{\!n}}}{{\hbar}}}\right) \\
&-\sum_{n=1}^\infty (-1)^{n+1} \frac{1}{2\pi}e^{-\tau V^\ddagger/\hbar}\mathrm{sgn}\!\left(n\tau_{\rm c}-\tau\right) \sqrt{\frac{-S_{\!n,\mathrm{inst}}''(\tau)}{2\Delta S_{\!n}}} \\
&+\sum_{n=1}^\infty (-1)^{n+1} \frac{1}{2\pi} e^{-\tau V^\ddagger/\hbar} \frac{1}{n\tau_{\rm c}-\tau}
\\
&+\sum_{n=0}^\infty (-1)^{n} \frac{1}{2\pi} e^{-\tau V^\ddagger/\hbar} \frac{1}{n\tau_{\rm c}+\tau}.
    \end{aligned}
\end{equation}
This can then be simplified by noting that the final two sums can be combined to give
\begin{equation}
    \sum_{n=1}^\infty (-1)^{n+1} \frac{1}{\frac{2\pi n}{\omega}-\tau} + \sum_{n=0}^\infty (-1)^{n} \frac{1}{\frac{2\pi n}{\omega}+\tau}=\sum_{n=-\infty}^\infty \frac{(-1)^n}{\tau+\frac{2\pi n}{\omega}}
\end{equation}
which is equivalent to 
\begin{equation}
    \sum_{n=-\infty}^\infty \frac{(-1)^n}{\tau+\frac{2\pi n}{\omega}} = \frac{\omega}{2\sin(\omega\tau/2)}.
\end{equation}
Hence, we arrive at our final expression for the uniform thermal rate in one dimension
\begin{equation}
    \begin{aligned}
        \!k Z_r&\sim \sum_{n=1}^\infty (-1)^{n+1} k_{n,\mathrm{inst}}Z_r\frac{1}{2}\mathrm{erfc}\!\left(\mathrm{sgn}\!\left(n\tau_{\rm c}-\tau\right)\!\sqrt{\frac{{\Delta S_{\!n}}}{{\hbar}}}\right) \\
&-\sum_{n=1}^\infty (-1)^{n+1} \frac{1}{2\pi}e^{-\tau V^\ddagger/\hbar}\mathrm{sgn}\!\left(n\tau_{\rm c}-\tau\right) \sqrt{\frac{-S_{\!n,\mathrm{inst}}''(\tau)}{2\Delta S_{\!n}}} \\
&+\frac{\omega}{4\pi\sin(\omega\tau/2)} e^{-\tau V^\ddagger/\hbar}. 
    \end{aligned} \label{eq:one-dimensional-result}
\end{equation}

\bibliography{references,extra_refs}

\end{document}

\maketitle

\renewcommand{\thepage}{S\arabic{page}}
\renewcommand{\theequation}{S\arabic{equation}}
\renewcommand{\thefigure}{S\arabic{figure}}
\renewcommand{\thetable}{S\arabic{table}}
\renewcommand{\thesection}{S\arabic{section}}
\renewcommand{\thesubsection}{S\arabic{section}.\arabic{subsection}}

\section{I. Behaviour of the action as $t\to\infty$}
In order to analyse the $t\to\infty$ behaviour of the action, $S(t)$, in Eq.~(24) of the main manuscript, we need to relate the time to the action. To do so we follow the approach given in Ref.~\citenum{ZinnJustin2021BookCh37}. This involves first relating the energy of the forward/reverse paths, $E_\pm$, to the time, $t_\pm=\pm t- i \tau/2$, and then using the fact that the Euclidean action is related to the energy according to
\begin{equation}
    \frac{\partial S_{\pm}}{\partial t_\pm} =  i E_\pm 
\end{equation}
to integrate to give the desired result.

To do so we begin by noting that the time for the forward path is given by
\begin{equation}
     t_{+} = t-i\tau/2 = \int_{x_r}^{x_p} \sqrt{\frac{1}{2(E_+-V(x))}} \mathrm{d}x. 
\end{equation}
where $x=\sqrt{m}q$ is the mass weighted position, and $x_r$ and $x_p$ are the mass weighted locations of the dividing surfaces.
To simplify notation we define the maximum of $V(x)$ to be at the origin, $V(x)=V^\ddagger-\frac{1}{2}\omega^2x^2$, and introduce the energy differences $\Delta E_+=E_+-V^\ddagger$ and $\Delta V(x)=V(x)-V^\ddagger$. Splitting the path into two halves around the maximum  we can thus write
\begin{equation}
    t_+ = \int_{x_r}^0 \sqrt{\frac{1}{2(\Delta E_+- \Delta V(x))}} \mathrm{d}x + \int_{0}^{x_p} \sqrt{\frac{1}{2(\Delta E_+- \Delta V(x))}} \mathrm{d}x.
\end{equation}
Now we consider the behaviour as $|\Delta E_+|\to 0$, taking just the second half of the path we can add and subtract the harmonic contribution to give 
\begin{equation}
\begin{aligned}
    \int_{0}^{x_p} \sqrt{\frac{1}{2(\Delta E_+- \Delta V(x))}} \mathrm{d}x &= \int_{0}^{x_p} \sqrt{\frac{1}{2\Delta E_+ +\omega^2 x^2}} \mathrm{d}x\\
     & + \int_{0}^{x_p} \sqrt{\frac{1}{2(\Delta E_+- \Delta V(x))}}  - \sqrt{\frac{1}{2\Delta E_++\omega^2 x^2}} \mathrm{d}x
\end{aligned}
\end{equation}
The harmonic part can then be evaluated analytically to give
\begin{equation}
    \int_{0}^{x_p} \sqrt{\frac{1}{2\Delta E_+ +\omega^2 x^2}} \mathrm{d}x = \frac{1}{2\omega} \ln(1+ \frac{x_p \omega\left(x_p\omega + \sqrt{2\Delta E_+ + x_p^2\omega^2}\right)}{\Delta E_+}) 
\end{equation}
where it is assumed that $x_p>0$.
which can be simplified in the limit as $|\Delta E_+|\to0$ to give
\begin{equation}
    \int_{0}^{x_p} \sqrt{\frac{1}{2\Delta E_+ +\omega^2 x^2}} \mathrm{d}x \sim  \frac{1}{2\omega} \ln(\frac{2x_p^2\omega^2}{\Delta E_+}).%
\end{equation}
As $|\Delta E_+|\to0$ it is clear that the remaining contribution to the forward path approaches a constant i.e.~
\begin{equation}
    \int_{0}^{x_p} \sqrt{\frac{1}{2(\Delta E_+- \Delta V(x))}}  - \sqrt{\frac{1}{2\Delta E_++\omega^2 x^2}} \mathrm{d}x \sim \int_{0}^{x_p} \sqrt{\frac{1}{2(- \Delta V(x))}}  - \sqrt{\frac{1}{\omega^2 x^2}} \mathrm{d}x
\end{equation}
Performing the same analysis on the integral from $x_r$ to $0$, and combining the results we obtain the following expression for $t_+$
\begin{equation}
    t_+ \sim \frac{1}{2\omega} \ln(\frac{2x_p^2\omega^2}{\Delta E_+}) + \frac{1}{2\omega} \ln(\frac{2x_r^2\omega^2}{\Delta E_+}) + \int_{x_r}^{x_p} \sqrt{\frac{1}{2(- \Delta V(x))}}  - \sqrt{\frac{1}{\omega^2 x^2}} \mathrm{d}x
\end{equation}
as $|\Delta E_+|\to0$. This can then be rearranged to give
\begin{equation}
    \Delta E_+ \sim 2 \omega^2 \sqrt{x_r^2 x_p^2}  \exp(-\omega t_+ +  \omega \int_{x_r}^{x_p} \sqrt{\frac{1}{2(- \Delta V(x))}}  - \sqrt{\frac{1}{\omega^2 x^2}} \mathrm{d}x ) 
\end{equation}
as $\Re(t_+)\to\infty$. Which we can relate back to the unshifted energy as 
\begin{equation}
    E_+ \sim V^\ddagger + \omega \frac{a(x_r,x_p)}{2}  \exp(-\omega t_+) 
\end{equation}
where 
\begin{equation}
    a(x_r,x_p)= 4 \omega \sqrt{x_r^2 x_p^2}  \exp(  \omega \int_{x_r}^{x_p} \sqrt{\frac{1}{2(- \Delta V(x))}}  - \sqrt{\frac{1}{\omega^2 x^2}} \mathrm{d}x ). 
\end{equation}

We can then also obtain the equivalent expression for $t_{-}=-t-i\tau/2$ by noting that 
\begin{equation}
    t_{-} = -t -i\tau/2 = \int_{x_p}^{x_r} \sqrt{\frac{1}{2(E_--V(x))}} \mathrm{d}x. 
\end{equation}
Then to evaluate the final result we just need to be careful to note that (because $x_r<0$)
\begin{equation}
    \int_{0}^{x_r} \sqrt{\frac{1}{2\Delta E_- +\omega^2 x^2}} \mathrm{d}x \sim  -\frac{1}{2\omega} \ln(\frac{2x_r^2\omega^2}{\Delta E_-})
\end{equation}
Hence, 
\begin{equation}
    t_- \sim -\frac{1}{2\omega} \ln(\frac{2x_p^2\omega^2}{\Delta E_-}) - \frac{1}{2\omega} \ln(\frac{2x_r^2\omega^2}{\Delta E_-}) + \int_{x_p}^{x_r} \sqrt{\frac{1}{2(- \Delta V(x))}}  - \sqrt{\frac{1}{\omega^2 x^2}} \mathrm{d}x,
\end{equation}
as $|\Delta E_-|\to0$.
Swapping the limits we see that $-t_-$ is related to $\Delta E_{-}$ as $t_+$ is to $\Delta E_+$ and hence 
\begin{equation}
    E_- \sim V^\ddagger + \omega \frac{a(x_r,x_p)}{2}  \exp(\omega t_-) 
\end{equation}
as $\Re(t_-)\to-\infty$.

Now we can combine the energies together and integrate to obtain an expression for the action
\begin{equation}
    S(t) = S_{+}(t_+) + S_{-}(t_-) = S_{\!\infty} + i\int_{\infty}^t E_{+}(t'-i\tau/2) - E_{-}(-t'-i\tau/2) \mathrm{d}t'.
\end{equation}
Inserting the asymptotic definitions of $E_{\pm}$ gives
\begin{equation}
    S(t) \sim S_{\!\infty} + i \omega \frac{a(x_r,x_p)}{2} \int_{\infty}^t e^{-\omega t'+i\tau\omega/2} - e^{-\omega t'-i\tau\omega/2} \mathrm{d}t'.
\end{equation}
Finally evaluating the integral and combining together the complex exponentials we obtain
\begin{equation}
    S(t) \sim S_{\!\infty} + a(x_r,x_p) \sin(\omega \tau/2) e^{-\omega t}
\end{equation}
as $t\to\infty$.

\section{II. Behaviour of the prefactor as $t\to\infty$}
The prefactor in Eq.~(24) of the main manuscript has the form
\begin{equation}
\begin{aligned}
    A(t)&=\frac{-1}{4m^2} \sqrt{\frac{C_+C_-}{(2\pi\hbar)^2}} \Bigg(\frac{\partial S_-}{\partial s_r}\frac{\partial S_+}{\partial s_p}-\frac{\partial S_-}{\partial s_p}\frac{\partial S_-}{\partial s_r}
    -\frac{\partial S_+}{\partial s_p}\frac{\partial S_+}{\partial s_r}+\frac{\partial S_-}{\partial s_p}\frac{\partial S_+}{\partial s_r}\Bigg)
    \end{aligned}
\end{equation}
where 
\begin{equation}
    C_{\pm} =  -\frac{\partial^2 S_\pm}{\partial s_r \partial s_p}.
\end{equation}
In order to obtain an expression for the prefactor of the correlation function as $t\to\infty$ we, therefore, need to consider the derivatives of the action with respect to the end points, making use of results from the previous section.

 To begin with we note that by defining
\begin{equation}
\begin{aligned}
    \Delta S_{\!\pm} &= \Delta S_{\pm}|_{t=\infty} + \int_{\pm\infty}^{t_{\pm}} i\Delta E_{\pm}(t'_{\pm}) \mathrm{d}t'_{\pm} \\ &=  \Delta S_{\pm}|_{t=\infty} \mp\frac{ia(x_r,x_p)}{2}\exp(\mp \omega t_{\pm}) \\&= \Delta S_{\pm}|_{t=\infty}  - \frac{a(x_r,x_p)}{2} e^{\pm i(\pi/2+\tau\omega/2)} e^{-\omega t}
    \end{aligned}
\end{equation}
it follows that, as $V^\ddagger$ is a position independent constant, the derivative are related to the full action as
\begin{equation}
   \frac{ \partial\Delta S_{\!\pm}}{\partial s_{r}} = \frac{ \partial S_{\!\pm}}{\partial s_{r}}
\end{equation}
and
\begin{equation}
   \frac{ \partial\Delta S_{\!\pm}}{\partial s_{p}} = \frac{ \partial S_{\!\pm}}{\partial s_{p}}.
\end{equation}

Hence, we need
\begin{equation}
    \frac{\partial a}{\partial x_r} = \frac{a}{x_r} - \omega \Bigg(\sqrt{\frac{1}{2(-\Delta V(x_r))}} - \sqrt{\frac{1}{\omega^2 x_r^2}} \Bigg)a = -a \omega \sqrt{\frac{1}{2|\Delta V(x_r)|}}
\end{equation}
and
\begin{equation}
    \frac{\partial a}{\partial x_p} = \frac{a}{x_p} + \omega \Bigg(\sqrt{\frac{1}{2(-\Delta V(x_p))}} - \sqrt{\frac{1}{\omega^2 x_p^2}} \Bigg) a = a \omega \sqrt{\frac{1}{2|\Delta V(x_p)|}}
\end{equation}
from which we also have
\begin{equation}
\begin{aligned}
    \frac{\partial^2 a}{\partial x_r \partial x_p} 
&=-a  \omega^2 \sqrt{\frac{1}{2|\Delta V(x_p)|}}    \sqrt{\frac{1}{2|\Delta V(x_r)|}} .
    \end{aligned}
\end{equation}

To arrive at the final result for the prefactor we need to consider
\begin{equation}
    \frac{\partial S_+}{\partial s_\alpha} = \sqrt{m}  \frac{\partial S_+}{\partial x_\alpha} \sim \sqrt{m}  \frac{\partial S_+}{\partial x_\alpha}\bigg|_{t=\infty} -\sqrt{m} \frac{1}{2} e^{i (\pi/2+\tau\omega/2) } e^{-\omega t} \frac{\partial a}{\partial x_\alpha}
\end{equation}
\begin{equation}
    \frac{\partial S_-}{\partial s_\alpha} =  \sqrt{m}  \frac{\partial S_-}{\partial x_\alpha} \sim \sqrt{m}  \frac{\partial S_-}{\partial x_\alpha}\bigg|_{t=\infty} - \sqrt{m} \frac{1}{2} e^{-i (\pi/2+\tau\omega/2) } e^{-\omega t} \frac{\partial a}{\partial x_\alpha}
\end{equation}
where using the connection between the derivative of the action and the (imaginary-time) momentum 
\begin{equation}
      \frac{\partial S_\pm}{\partial s_r} = \mp \tilde{p}(s_r,E)= \mp  \sqrt{2m(V(s_r)-E)}
\end{equation}
\begin{equation}
      \frac{\partial S_\pm}{\partial s_p} = \pm \tilde{p}(s_p,E) = \pm \sqrt{2m(V(s_p)-E)}
\end{equation}
we have that
\begin{equation}
      \frac{\partial S_\pm}{\partial s_r}\bigg|_{t=\infty} = \mp  \sqrt{2m(V(s_r)-V^\ddagger)} = \mp i \sqrt{2m|\Delta V(s_r)|}
\end{equation}
\begin{equation}
      \frac{\partial S_\pm}{\partial s_p}\bigg|_{t=\infty}  = \pm \sqrt{2m(V(s_p)-V^\ddagger)} = \pm i \sqrt{2m|\Delta V(s_p)|}.
\end{equation}
Differentiating once more we obtain
\begin{equation}
    \frac{\partial^2 S_+}{\partial s_r \partial s_p}  \sim -{m} \frac{1}{2} e^{i (\pi/2+\tau\omega/2) } e^{-\omega t} \frac{\partial^2 a}{\partial x_r \partial x_p}
\end{equation}
\begin{equation}
    \frac{\partial^2 S_-}{\partial s_r \partial s_p}  \sim -{m} \frac{1}{2} e^{-i (\pi/2+\tau\omega/2) } e^{-\omega t} \frac{\partial^2 a}{\partial x_r \partial x_p}.
\end{equation}
Combining these results we can evaluate asymptotic behaviour of the part of the prefactor that involves derivatives of the action
\begin{equation}
\begin{aligned}
    &\sqrt{\frac{\partial^2 S_+}{\partial s_r \partial s_p} \frac{\partial^2 S_-}{\partial s_r \partial s_p}}\left(\frac{\partial S_-}{\partial s_r}\frac{\partial S_+}{\partial s_p}-\frac{\partial S_-}{\partial s_p}\frac{\partial S_-}{\partial s_r}-\frac{\partial S_+}{\partial s_p}\frac{\partial S_+}{\partial s_r}+\frac{\partial S_-}{\partial s_p}\frac{\partial S_+}{\partial s_r}\right) \\
& \sim  \sqrt{\frac{m^2}{4}e^{-2\omega t}\left(\frac{\partial^2 a}{\partial x_r \partial x_p}\right)^2} \left(4i^2 \sqrt{2m|\Delta V(x_r)|}\sqrt{2m|\Delta V(x_p)|}  \right) \\
& = \frac{m}{2} e^{-\omega t} a \omega^2 \frac{1}{\sqrt{2|\Delta V(x_p)|}} \frac{1}{\sqrt{2|\Delta V(x_r)|}} \left(4i^2 \sqrt{2m|\Delta V(x_r)|}\sqrt{2m|\Delta V(x_p)|}  \right) \\
& = -2 m^2 \omega^2 e^{-\omega t} a(x_r,x_p) 
    \end{aligned}
\end{equation}
Hence it follows that
\begin{equation}
\begin{aligned}
    A(t)&=\frac{-1}{4m^2} \sqrt{\frac{C_+C_-}{(2\pi\hbar)^2}} \Bigg(\frac{\partial S_-}{\partial s_r}\frac{\partial S_+}{\partial s_p}-\frac{\partial S_-}{\partial s_p}\frac{\partial S_-}{\partial s_r}
    -\frac{\partial S_+}{\partial s_p}\frac{\partial S_+}{\partial s_r}+\frac{\partial S_-}{\partial s_p}\frac{\partial S_+}{\partial s_r}\Bigg)\\
&=\frac{1}{8m^2\pi\hbar}  2 m^2 \omega^2 e^{-\omega t} a(x_r,x_p) = \frac{a \omega^2}{4\pi\hbar}   e^{-\omega t} .
    \end{aligned}
\end{equation}

\section{III. Behaviour of Upadhyayula and Pollak's theory as $\hbar\to0$}
Recently, Upadhyayula and Pollak have suggested a theory that they name ``uniform semiclassical instanton rate theory''. Here we clarify that, while their theory is based on the uniform semiclassical transmission probability of Kemble,\cite{Kemble1935WKB} it is not a rigorous uniform semiclassical theory in $\tau$ as $\hbar\to0$. It is important to stress that Upadhyayula and Pollak do not explicitly claim their theory to be uniform in $\tau$ as $\hbar\to0$. Nevertheless, for the general reader confusion about this issue may arise, particularly given they derive their result starting from a rigorous uniform semiclassical theory in $E$ as $\hbar\to0$.

Upadhyayula and Pollak's theory can be written in one dimension as
\begin{equation}
    \begin{aligned}
k_{\text{Ref.~\citenum{Upadhyayula2023UniformInstanton}}}Z_r = \frac{e^{-\Phi_\tau(\mathcal{E}_\tau)/\hbar}}{2\pi\tau}  &\left[ \sqrt{\frac{\pi\tau^2/2\hbar}{ \Phi_\tau''(\mathcal{E}_\tau)}}\mathrm{erfc}\left(-\sqrt{\frac{\tau^2/2\hbar}{ \Phi_\tau''(\mathcal{E}_\tau)}}\right)\right.+\left.\exp(-\frac{\tau^2}{2\hbar \Phi_\tau''(\mathcal{E}_\tau)})\vphantom{\sqrt{\frac{\tau^2/2\hbar}{ \Phi_\tau''(\mathcal{E}_\tau)}}}\right] 
\end{aligned}
\end{equation}
where the effective action $\Phi_\tau(E)$ is defined as
\begin{equation}
    \Phi_\tau(E) = \tau E +\hbar \ln\left( 1+e^{W(E)/\hbar}\right)
\end{equation}
and the stationary energy, $\mathcal{E}_\tau$, is given by 
\begin{equation}
    \Phi_\tau'(\mathcal{E}_\tau) = \tau +\frac{W'(\mathcal{E}_\tau)}{1+e^{-W(\mathcal{E}_\tau)/\hbar}} = 0.
\end{equation}
Here, $W(E)$ is the reduced action (the Legendre transform of $S_{\!\rm inst}(\tau)$). Note that if $W(E)$ is monotonic and $\tau(E)$ is monotonic then it follows that $\tau$ is also a monotonic function of $\mathcal{E}_\tau$  
\begin{equation}
    \tau = -\frac{W'(\mathcal{E}_\tau)}{1+e^{-W(\mathcal{E}_\tau)/\hbar}}.
\end{equation}

To illustrate the asymptotic behaviour of Upadhyayula and Pollak's theory we consider the asymmetric Eckart barrier (defined in main text) at various values of $\hbar$. With the exception of $\hbar$, the parameters of the model are defined to be the same in reduced units as those in the main text with $m=1$, $L=8/\sqrt{3\pi}$, 
$V_1=(6/\pi)^{\frac{1}{4}}$, 
$V_2=(24/\pi)^{\frac{1}{4}}$,  
and $\hbar$ varying from $1$ to $1/8$. Note that varying $\hbar$ while keeping the other constants fixed simply corresponds to a change in the definition of the reduced units. An asymptotic theory should obey,
\begin{equation}
    \lim_{\hbar\to0}\frac{k_{\rm approx}}{k_{\rm exact}} = 1,
\end{equation} 
and a uniform asymptotic theory should obey this at every value of the uniform parameter (in this case $\tau$). Hence, by plotting $\frac{k_{\rm approx}}{k_{\rm exact}}$ as a function of $\tau$ for successively smaller values of $\hbar$ we can test whether a theory is uniformly asymptotic.

\begin{figure}[t]
    \centering
    \includegraphics[width=0.5\linewidth]{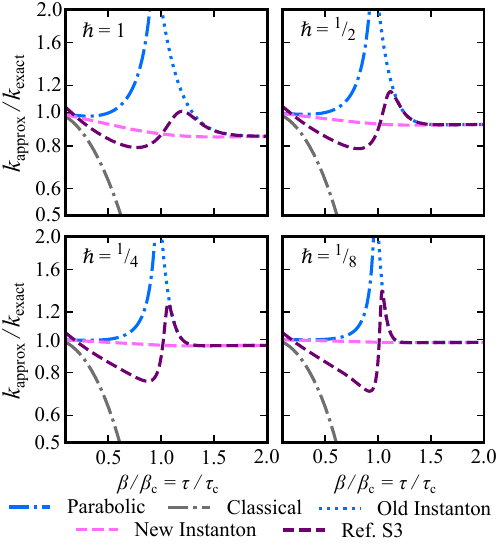}
    \caption{Plot showing the errors for various approximate methods for the asymmetric Eckart barrier as a function of inverse temperature, for decreasing values of $\hbar$. This illustrates that while the new instanton theory presented in this work is a rigorous semiclassical theory that is uniform in $\tau$ , the theory of Ref.~\citenum{Upadhyayula2023UniformInstanton} is not. Note that the errors are plotted on a logarithmic scale.}
    \label{fig:hbar_plot_asym_eckart}
\end{figure}

Figure \ref{fig:hbar_plot_asym_eckart} shows the ratio of the rate constant calculated using various approximate theories to the exact result at four different values of $\hbar$ as a function of $\tau/\tau_{\rm c}$ for this model. The parameters considered in the top left panel, where $\hbar=1$, are equivalent to Fig.~5 of the main text and also to those considered in Ref.~\citenum{Upadhyayula2023UniformInstanton}. In this regime the numerical deviation between the theory of Ref.~\citenum{Upadhyayula2023UniformInstanton} and the new theory presented here is relatively small. However, it is notable that approaching the crossover temperature from below $k_{\text{Ref.~\citenum{Upadhyayula2023UniformInstanton}}}$ initially tracks the increase in the instanton rate, before oscillating to underestimate the rate just above the crossover temperature. This behaviour is exacerbated as $\hbar$ decreases, with $k_{\text{Ref.~\citenum{Upadhyayula2023UniformInstanton}}}$ following $k_{\text{inst}}$ more and more closely below the crossover temperature, and the oscillation becoming more and more pronounced. In contrast, the ratio between the exact result and the theory presented in this work approaches 1 as $\hbar\to0$ for all values of $\tau/\tau_{\rm c}$ as is expected for a uniform theory. %

\clearpage
\bibliography{references,extra_refs} %